\documentclass[11pt,a4paper]{article}
\pdfoutput=1
\usepackage{jheppub}
\usepackage{rotating}
\usepackage{array}
\usepackage{amsmath}
\usepackage[normalem]{ulem}
\usepackage{slashed}
\usepackage{booktabs}
\usepackage[pdftex,table,dvipsnames]{xcolor}
\usepackage{units}
\usepackage{multirow}
\usepackage{url}

\usepackage{xspace}
\usepackage{subfig}
\usepackage{color}

\def \belletwo {Belle\,II\xspace}
\def \belle {Belle\xspace}
\def \babar {BaBar\xspace}
\def \superkekb {SuperKEKB\xspace}
\def \bfactory {\textit{B}-factory\xspace}
\def \bfactories {\textit{B}-factories\xspace}

\preprint{DESY-19-141, OUTP-19-10P, TTK-19-46}

\title{Invisible and displaced dark matter signatures at \belletwo}

\author[1]{Michael Duerr,}
\author[2]{Torben Ferber,}
\author[3]{Christopher Hearty,}
\author[4]{Felix Kahlhoefer,}
\author[2]{Kai Schmidt-Hoberg,}
\author[4]{and Patrick Tunney}

\affiliation[1]{ Rudolf Peierls Centre for Theoretical Physics, University of Oxford, Clarendon Laboratory, Parks Road, Oxford OX1 3PU, United Kingdom}
\affiliation[2]{ DESY, Notkestrasse 85, D-22607 Hamburg, Germany}
\affiliation[3]{ Department of Physics and Astronomy, University of British Columbia, Vancouver, British Columbia, Canada V6T 1Z1; also at Institute of Particle Physics (Canada) }
\affiliation[4]{ Institute for Theoretical Particle Physics and Cosmology (TTK), RWTH Aachen University, D-52056 Aachen, Germany}

\emailAdd{michael.duerr@physics.ox.ac.uk}
\emailAdd{torben.ferber@desy.de}
\emailAdd{hearty@physics.ubc.ca}
\emailAdd{kahlhoefer@physik.rwth-aachen.de}
\emailAdd{kai.schmidt.hoberg@desy.de}
\emailAdd{tunney@physik.rwth-aachen.de}

\abstract{
Many dark matter models generically predict invisible and displaced signatures at \belletwo, but even striking events may be missed by the currently implemented search programme because of inefficient trigger algorithms. Of particular interest are final states with a single photon accompanied by missing energy and a displaced pair of electrons, muons, or hadrons. We argue that a displaced vertex trigger will be essential to achieve optimal sensitivity at \belletwo. To illustrate this point, we study a simple but well-motivated model of thermal inelastic dark matter in which this signature naturally occurs and show that otherwise inaccessible regions of parameter space can be tested with such a search. We also evaluate the sensitivity of single-photon searches at \babar and \belletwo to this model and provide detailed calculations of the relic density target.
}

\keywords{Mostly Weak Interactions: Beyond Standard Model; Collider Physics: $e^+$-$e^-$ Experiments; Astroparticles: Cosmology of Theories beyond the SM}

\begin{document}

\maketitle

\flushbottom

\section{Introduction}

While there is strong evidence for the existence of dark matter~(DM) over a very large range of astrophysical scales,
no clear sign of its particle physics nature has been established to date. 
If DM has non-negligible couplings to Standard Model~(SM) states it could potentially be produced at particle colliders or be observed in direct and indirect detection experiments.
For DM masses above a few GeV, direct detection experiments in particular have put very severe bounds on the 
DM scattering cross section~\cite{Aprile:2018dbl,Ren:2018gyx}, while smaller DM masses are less constrained because of the finite threshold energy required by these experiments.
Correspondingly a lot of attention is currently focused on rather light DM and associated dark sector states with masses in the MeV to GeV range~\cite{Batell:2009di,Andreas:2012mt,Schmidt-Hoberg:2013hba,Essig:2013vha,Izaguirre:2013uxa,Batell:2014mga,Dolan:2014ska,Krnjaic:2015mbs,Dolan:2017osp,Izaguirre:2017bqb,Knapen:2017xzo,Beacham:2019nyx,Bernreuther:2019pfb,Bondarenko:2019vrb,Filimonova:2019tuy}. 
For such light dark sectors the couplings to SM states are constrained to be rather small and high-energy machines such as the LHC are not
necessarily the most promising tools to explore such scenarios. In fact low-energy but high-intensity facilities such as \bfactories have unique advantages.

In this work we explore the sensitivity of the \belletwo detector to light dark sectors with a particular focus on signatures that may be missed with the current experimental configuration.
While the primary purpose of \belletwo is to study the properties of $B$-mesons~\cite{Abashian:2000cg,Abe:2010gxa}, its hermetic detector and optimized triggers also allow for searches for various DM models. The simplest signature of direct DM production at  \belletwo is an excess of events with 
a single high-energy photon and a large amount of missing energy, which is well established and integrated into the current search program~\cite{Kou:2018nap}. 
Our main focus will be on another key signature which generically appears in a number of models, consisting of a single photon accompanied by missing energy and a displaced pair of electrons, muons, or hadrons.

Such a signature arises for example if DM interactions involve an inelastic transition between two states $\chi_1$ and $\chi_2$ with small mass splitting. This interesting scenario allows for light DM production via thermal freeze out consistent with constraints from the Cosmic Microwave Background (CMB) and direct detection experiments. In this work we will consider a simple inelastic DM model in order to study the signature 
in detail, noting that the same final state could e.g.\ also arise in scenarios with strongly coupled DM~\cite{Berlin:2018tvf}.
As the signature we consider is vetoed in the mono-photon analysis because of the presence of additional final state particles, it requires a new search strategy that has not been performed at a collider yet.
Indeed, we find that it is crucial to develop new trigger algorithms, in particular a new displaced vertex trigger, in order to fully exploit the potential of  \belletwo to uncover the nature of DM.

The paper is organized as follows. Section \ref{SEC:theory} provides an overview of the inelastic DM model and the calculation of relic targets. 
Section \ref{SEC:belletwo} describes the recast of the \babar  and \belletwo mono-photon searches~\cite{Lees:2017lec,Kou:2018nap}, and describes the calculation of the sensitivity to the new displaced signature.
Finally, in section \ref{SEC:discussion}, we compare the \belletwo sensitivities with other existing constraints from beam dumps and expected sensitivities from long-lived particles searches at CERN.

\section{Inelastic dark matter}
\label{SEC:theory}

\subsection{Light thermal dark matter}
A particularly appealing scenario for the production of DM is thermal freeze out, which is insensitive to the initial conditions
of the early universe and therefore very predictive. This mechanism requires significant couplings to SM states
to allow both for an initial thermalisation of the dark sector as well as sufficiently effective annihilations to be consistent 
with the observed relic abundance $\Omega h^2 = 0.12$~\cite{Aghanim:2018eyx}. While DM particles with mass below a few GeV are less constrained by direct searches as discussed above, there are strong constraints on late-time annihilations from observations of
the CMB anisotropies, basically ruling out the case of elastic scattering if the 
annihilations proceed via $s$-wave~\cite{Aghanim:2018eyx}.\footnote{In fact thermal DM is also strongly constrained by the 
requirement of successful primordial nucleosynthesis, excluding $m_\chi \lesssim$ \unit[10]{MeV} with a slight dependence on the quantum numbers of DM, see e.g.~\cite{Depta:2019lbe} for a recent evaluation of the corresponding bounds.}

An interesting and simple idea to reconcile the light DM case with these constraints is to assume a small mass splitting
between two DM states $\chi_1$ and $\chi_2$ which are coupled off-diagonally to a new mediator. The dominant annihilation channel will then be 
coannihilations,  $\chi_1 \chi_2 \rightarrow \text{SM}$. If the heavier state $\chi_2$ is unstable with a sufficiently short lifetime,
no $\chi_2$ particles will be available during recombination, such that the main annihilation channel is no longer 
active and CMB bounds are evaded. In addition bounds from direct detection experiments are further diminished or even absent,
as inelastic scatterings are suppressed kinematically and elastic scatterings have a loop-suppressed cross section.
The only way to conclusively test this scenario is therefore via accelerator experiments. Given that this set-up is a well-motivated scenario for 
light thermal DM, it has previously been discussed in the literature in the context of particle colliders~\cite{Izaguirre:2015zva,Berlin:2018jbm}, fixed-target experiments~\cite{Izaguirre:2017bqb} and the muon anomalous magnetic moment~\cite{Mohlabeng:2019vrz}. 

\subsection{A simple model}
\label{SEC:theory_model}

Let us consider a dark sector fermion $\psi = \psi_L + \psi_R$  charged under a dark gauge group $U(1)_X$ but singlet under the SM gauge group.
In addition we assume the presence of a dark sector scalar $\phi$ with trilinear couplings to dark fermion bilinears (i.e.\ $\phi$ is charged under the $U(1)_X$ gauge symmetry), which will generate a Majorana mass term after spontaneously breaking $U(1)_X$. Note that a Dirac mass $m_D$ is gauge-invariant and hence independent of spontaneous symmetry breaking.
The Lagrangian is then given by
\begin{equation}\label{eq:Lagrangian}
 \mathcal{L}_\psi = i \overline{\psi} \slashed{D} \psi - m_D \overline{\psi} \psi - \lambda_1 \phi  \overline{\psi_L^c} \psi_L - \lambda_2 \phi  \overline{\psi_R^c} \psi_R - V(\phi) + \text{h.c.},
\end{equation}
where  $V(\phi)$ is the scalar potential for $\phi$. We assume that this potential leads to a vacuum expectation value of $\phi$ which will then generally provide a Majorana mass for the left- and right-handed part of $\psi$.\footnote{The same scalar may also give a mass to the gauge boson $\hat{X}_\mu$, but we will refrain from making detailed assumptions regarding the mass generation of the gauge boson.}  This results in the following mass matrix,
\begin{equation}
 \mathcal{L}_\psi \supset -\frac{1}{2} \begin{pmatrix} \overline{\psi_L^c} & \overline{\psi_R} \end{pmatrix} \begin{pmatrix} m_L & m_D \\ m_D & m_R \end{pmatrix} \begin{pmatrix}
                                                                                                                                                                                               \psi_L \\ \psi_R^c
                                                                                                                                                                                              \end{pmatrix}
                                                                                                                                                                                       + \text{h.c.} \;,
\end{equation}
which can be diagonalized with the mixing matrix
\begin{equation}
 \mathcal{U} = \begin{pmatrix}
                \cos \theta & \sin \theta \\
                i \sin \theta & -i \cos\theta
               \end{pmatrix} \; .
\end{equation}
We denote the corresponding (Majorana) mass eigenstates by $\chi_1$ and $\chi_2$, and the relation to the left- and right-handed components of $\psi$ is given by 
\begin{align}
 \psi_L &= \cos \theta \, \chi_{1,L} + i \sin \theta  \, \chi_{2,L} , \\
 \psi_R &= \sin \theta  \, \chi_{1,R} + i \cos \theta  \, \chi_{2,R} .
\end{align}
Rewriting the Lagrangian for $\psi$ in terms of $\chi_1$ and $\chi_2$ with masses $m_{\chi_1}$ and $m_{\chi_2}$ under the assumption that $m_L = m_R$ (which means $\lambda_1 = \lambda_2$ in Eq.~\eqref{eq:Lagrangian} above), we obtain a purely off-diagonal coupling of the DM states to the $U(1)_X$ gauge boson $\hat{X}$:\footnote{Note that for $\lambda_1 \neq \lambda_2$ the coupling is still dominantly off-diagonal if $m_D \gg m_L, m_R$~\cite{Izaguirre:2015zva}.}
\begin{equation}\label{eq:Lpsi}
\mathcal{L}_\psi = i \overline{\chi_1} \slashed{\partial} \chi_1 + i \overline{\chi_2} \slashed{\partial} \chi_2 + \frac{i}{2} g_X \hat{X}_\mu \overline{\chi_2} \gamma^\mu \chi_1 - \frac{i}{2} g_X \hat{X}_\mu \overline{\chi_1} \gamma^\mu \chi_2 -\frac{1}{2} m_{\chi_1} \overline{\chi_1} \chi_1  -\frac{1}{2} m_{\chi_2} \overline{\chi_2} \chi_2. 
\end{equation}
Here, $g_X$ is the gauge coupling of $U(1)_X$, and we have assumed that $\psi$ has a $U(1)_X$ charge of unity. 

\subsection{Couplings to the Standard Model}
In order to fully define the set-up we have to specify the couplings to SM states.
Potential renormalisable inter-sector couplings which are allowed by the gauge symmetry correspond to kinetic mixing of the new gauge boson $\hat{X}$ with the SM hypercharge gauge boson $Y$, or to a mixing of the dark sector scalar $\phi$ with the SM Higgs boson $H$.
In general both couplings are expected to be present, resulting in a `two mediator' model with a rather complex phenomenology as discussed e.g.\ in Refs.~\cite{Duerr:2016tmh,Darme:2017glc}.
For the present discussion we will assume that the dominant interaction is generated by kinetic mixing and neglect a potential scalar mixing.\footnote{In fact, for the mass range we are going to consider, the scalar portal
will not be able to accommodate the measured relic abundance while being compatible with experimental constraints due to the Yukawa suppression of couplings to light SM fermions.}

The most general renormalisable Lagrangian for the SM with a new $U(1)_X$ gauge boson $\hat{X}$ with mass $m_{\hat{X}}$ is given by
\begin{equation}\label{eq:Lkinetic}
 \mathcal{L} = \mathcal{L}_\text{SM} - \frac{1}{4} \hat{X}_{\mu\nu} \hat{X}^{\mu\nu} + \frac{1}{2} m_{\hat{X}}^2 \hat{X}_{\mu} \hat{X}^{\mu} -\frac{\epsilon}{2 c_\text{W}}  \hat{X}_{\mu\nu} \hat{B}^{\mu\nu} ,
\end{equation}
where the SM Lagrangian contains
\begin{equation}
 \mathcal{L}_\text{SM} \supset - \frac{1}{4} \hat{B}_{\mu\nu} \hat{B}^{\mu\nu} - \frac{1}{4} \hat{W}^a_{\mu\nu} \hat{W}^{a\mu\nu} + \frac{1}{2} m_{\hat{Z}}^2 \hat{Z}_{\mu} \hat{Z}^{\mu}. 
\end{equation}

We denote the gauge fields (and the corresponding masses) in the original basis before diagonalisation by hats, such that $\hat{B}_{\mu\nu}$, $\hat{W}_{\mu\nu}$, and $\hat{X}_{\mu\nu}$ are the field strength tensors of $U(1)_Y$, $SU(2)_L$, and $U(1)_X$, respectively. We choose to normalize the kinetic mixing parameter $\epsilon$ by the (physical) value of the cosine of the Weinberg angle $c_W$ such that the coupling to electromagnetism is given by 
$\tfrac{\epsilon}{2}  \hat{X}^{\mu\nu} \hat{F}^\text{em}_{\mu\nu}$ to match the usual notation of the kinetic mixing term in the dark photon literature. We assume that there is no mass mixing between $\hat{X}$ and $\hat{Z}$, which could arise if either the SM Higgs is charged under $U(1)_X$ or the new scalar field is charged under both the SM gauge group and $U(1)_X$. 

The field strengths are diagonalized and canonically normalized by two consecutive transformations, to connect the original (hatted) fields to the physical photon $A_\mu$, the physical $Z$-boson $Z_\mu$, and the new physical gauge boson $A^\prime_\mu$ with mass $m_{A^\prime}$ (`dark photon'), as discussed in detail in Refs.~\cite{Babu:1997st,Frandsen:2011cg}.

The free parameters of the model are then $m_{A^\prime}$, $m_{\chi_1}$, $\Delta = m_{\chi_2} - m_{\chi_1}$, $\epsilon$, and $\alpha_\text{D} = g_X^2/ 4 \pi$. Here we will concentrate on the case $m_{A^\prime} > m_{\chi_1}+m_{\chi_2}$ such that the decay $A^\prime \rightarrow \chi_1 \chi_2$ is kinematically allowed and hence the dominant decay channel.\footnote{For 
$m_{A^\prime} < m_{\chi_1}+m_{\chi_2}$ the dark photon has to decay to SM states, a scenario which is covered by a large number of searches.}

\subsection{Relic density and thermal targets}

To put constraints on light DM into context, it is useful to identify a thermal target, i.e.\ a region in parameter space in which the measured DM relic abundance
is reproduced by thermal freeze out. The requirement $m_{A'} > m_{\chi_1} + m_{\chi_2}$ ensures that the annihilation channel $\chi_1 \chi_1 \rightarrow A^\prime A^\prime$ is closed, which is crucial for the scenario to be viable. The reason is that these annihilations, which would proceed via a $t$-channel exchange of $\chi_2$, would still be active during recombination and the subsequent $A^\prime$ decays into SM particles would lead to unacceptably large energy injection.\footnote{As a light $A^\prime$ basically couples to charge, for $m_{A^\prime} \lesssim 1$~MeV the dark photon would be very long-lived as decays into electron-positron pairs are no longer kinematically available. However, also this parameter region is excluded cosmologically due to the extra energy density stored in the  $A^\prime$ particles (see e.g.~\cite{Bringmann:2016din}).}
 
For the assumed mass hierarchy the DM freeze out is instead dominated by the coannihilation channel  $\chi_1 \chi_2 \rightarrow A^{\prime \ast} \rightarrow \text{SM}$,
which is no longer active during recombination as the $\chi_2$ abundance is negligible. 
To leading order in the relative velocity $v$ the corresponding $s$-wave co-annihilation cross section can be written as~\cite{Izaguirre:2015zva} 
\begin{equation}
\sigma v (\chi_1 \chi_2 \to e^+ e^-) \approx  \frac{4 \pi \epsilon^2 \alpha \alpha_\text{D} (m_{\chi_1}+m_{\chi_2})^2}{[(m_{\chi_1}+m_{\chi_2})^2-m_{A^\prime}^2]^2+m_{A^\prime}^2 \Gamma_{A^\prime}^2}\; .
\end{equation}
Here, $\alpha$ is the fine-structure constant and $\Gamma_{A^\prime}$ is the width of the dark photon. 
Crucially the relic abundance is set by the product of dark and visible couplings, $\epsilon^2 \alpha_\text{D}$, for this annihilation channel (rather than by the dark coupling alone), such that thermal freeze out can
be constrained by searches sensitive to the visible coupling $\epsilon$. 
Specifically, requiring the dark coupling to remain perturbative while constraining the visible coupling from above with particle physics experiments will allow to test the thermal freeze out conclusively. 

The annihilation cross section required to reproduce the observed DM relic abundance (and therefore the associated couplings) must in general be larger than in the elastic case in order to compensate the additional Boltzmann suppression of the coannihilation partner $\chi_2$. For large mass splittings $\Delta$ the required couplings are typically in conflict with existing experimental limits from LEP (see below) and we therefore limit ourselves to \hbox{$\Delta < 0.5 \,m_{\chi_1}$}. On the other hand, for mass splittings smaller than twice the electron mass, \hbox{$\Delta \lesssim \unit[1]{MeV}$}, the lifetime of the heavier state $\chi_2$ becomes so long that it violates cosmological bounds. 
Furthermore the thermal freeze out paradigm is in conflict with Big Bang Nucleosynthesis for DM masses below \hbox{$m_{\chi_1} \lesssim \unit[10]{MeV}$}~\cite{Boehm:2012gr,Depta:2019lbe}, so that we concentrate on \hbox{$m_{\chi_1} \ge \unit[10]{MeV}$}. 

To calculate the relic abundance in this model, we employ \texttt{micrOMEGAs v5.0.6}~\cite{Belanger:2018mqt}, using a \texttt{CalcHEP} model file~\cite{Belyaev:2012qa} implemented via \texttt{FeynRules v2.3.32}~\cite{Alloul:2013bka}. 
To take into account the effects from hadronic resonances mixing with the photon, we follow the usual approach, see e.g.~\cite{Izaguirre:2015yja,Berlin:2018bsc}, and calculate the annihilation cross section for $\chi_1 \chi_2 \to \mu^+ \mu^-$ analytically and rescale it with the measured value of $R(s) \equiv \sigma(e^+ e^- \rightarrow \text{hadrons})/\sigma(e^+ e^- \rightarrow \mu^+ \mu^-)$~\cite{Ezhela:2003pp,Tanabashi:2018oca} to obtain the cross section for $\chi_1 \chi_2 \to \text{hadrons}$. While the broader hadronic resonances are resolved in the $R(s)$ data, narrower ones like $J/\psi$ are not visible. As these resonances may have a sizeable impact on the calculated thermal target, we perform the needed thermal average analytically and correct the numerical result accordingly.

\begin{figure}[tbp]
\centering
 \includegraphics[scale=.727]{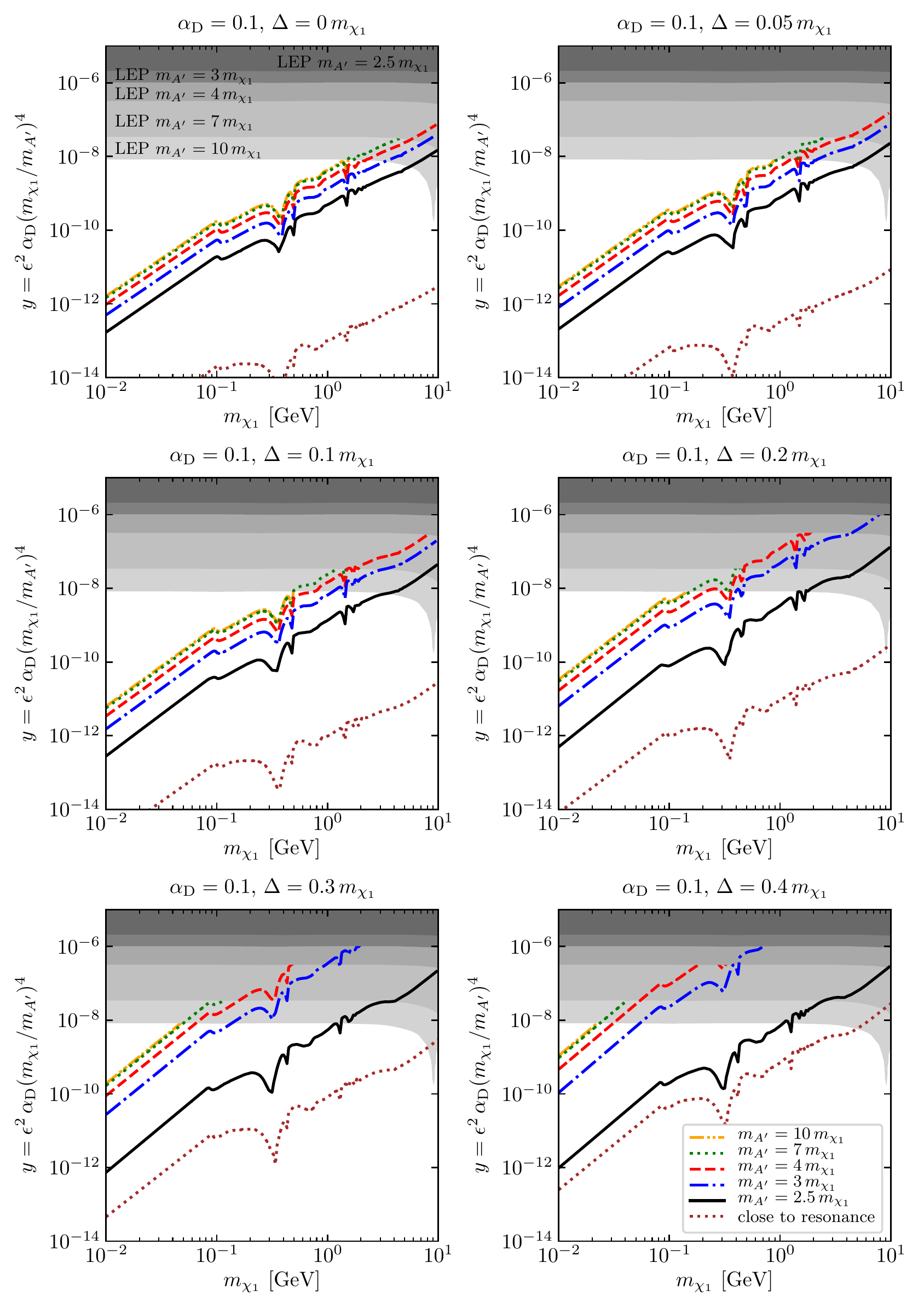}
 \caption{Thermal targets ($\Omega_\text{DM} h^2 = 0.12$~\cite{Aghanim:2018eyx}) for the inelastic DM model. We show various DM mass differences $\Delta/m_{\chi_1} = [0, 0.05, 0.1, 0.2, 0.3, 0.4]$ (different panels) and various mediator mass ratios $m_{A^\prime}/m_{\chi_1} = [2.5, 3, 4, 7, 10]$ (different lines in each panel). For the line labelled `close to resonance' the mediator mass is set to $m_{A^\prime}= 2.01 \, m_{\chi_1} + \Delta$.  The model-independent LEP bound~\cite{Hook:2010tw} on the kinetic mixing parameter $\epsilon$ constrains values of $\epsilon \approx 3 \times 10^{-2}$ away from the $Z$ resonance and hence results in a different limit on $y$ for differing ratios of $m_{A^\prime} / m_{\chi_1}$. }\label{fig:relic}
\end{figure}

For $m_{A^\prime} \gg m_{\chi_1},m_{\chi_2} \gg \Delta$ (and neglecting the width of the dark photon) the relic density only depends on the commonly used dimensionless variable $y$, defined via
\begin{equation}
 \epsilon^2 \alpha_\text{D} \frac{m_{\chi_1}^4}{m_{A^\prime}^4} \frac{1}{m_{\chi_1}^2} \equiv \frac{y}{m_{\chi_1}^2} \; .
\end{equation}
The virtue of this parameterisation is that if one calculates the relic target in terms of $y$ as a function of $m_{\chi_1}$, the result applies irrespective of the relative sizes of $\alpha_\text{D},\epsilon,m_{\chi_1}/m_{A^\prime}$. We show the required value of $y$ to obtain the measured relic density for various choices of the other parameters in figure~\ref{fig:relic}.\footnote{We provide tabulated values of $m_{\chi_1}$ and $\epsilon$ that satisfy the relic density target, for various choices of $m_{A^\prime}/m_{\chi_1}$ and $\Delta/m_{\chi_1}$, in the ancillary files of the arXiv entry and the supplementary material of the published version.} We use $m_{\chi_1}$ and $\Delta = m_{\chi_2} - m_{\chi_1}$ to parametrize the DM masses throughout this paper. 

The model-independent LEP bound~\cite{Hook:2010tw} on the kinetic mixing $\epsilon$ is used to constrain the allowed values of $y$ from above. Note that, since this bound has a constant value of $\epsilon \approx 3 \times 10^{-2}$ away from the $Z$ resonance, it results in a bound on $y$ that depends on the mass ratio $m_{A^\prime} / m_{\chi_1}$, hence each of the shaded grey region corresponds to one of the thermal target curves with the same mass ratio in each panel.

We observe that the assumption that the relic density depends only on $y$ is strictly only true for $m_{A^\prime} \gg (m_{\chi_1} + m_{\chi_2})$, as one can see that the curves for $m_{A^\prime} = 7 \, m_{\chi_1}$ and $m_{A^\prime} = 10 \, m_{\chi_1}$ coincide in all panels, independent of the value of $\Delta$. For the value $m_{A^\prime} = 3 \, m_{\chi_1}$ typically used in the literature, there is up to an order of magnitude difference between the value of $y$ obtained for $m_{A^\prime} = 3 \, m_{\chi_1}$ and the value of $y$ obtained for a sufficiently large value of the mass ratio (such that $y$ is independent of the mass ratio once again).
The DM annihilation is resonantly enhanced for $m_{A^\prime} \approx 2 \, m_{\chi_1} + \Delta $ ($ = m_{\chi_1} + m_{\chi_2}$). Note the larger spread between the curves for larger values of the mediator mass ratio (across different panels of figure~\ref{fig:relic}). This is due to the fact that the resonance condition is satisfied earlier when reducing the mediator mass ratio for larger values of $\Delta$. Close to the resonance extremely small values of the kinetic mixing can be compatible with the observed relic abundance~\cite{Feng:2017drg}.

Various calculations of the thermal target for inelastic DM can be found in the literature~\cite{Izaguirre:2015yja,Izaguirre:2015zva,Izaguirre:2017bqb,Berlin:2018bsc,Berlin:2018jbm,Mohlabeng:2019vrz,Battaglieri:2017aum}, either using the standard semi-analytical approach~\cite{Gondolo:1990dk,Griest:1990kh,Edsjo:1997bg}, or relying on numerical tools such as \texttt{micrOMEGAs}~\cite{Belanger:2018mqt}. We find differences to some of the results and also note that the various calculations do not completely agree. Most importantly, the hadronic resonances have a much less dramatic effect on the thermal target curve than is suggested in some of the literature. In addition, there appears to be some offset in overall normalisation between the different results. Our results reproduce the thermal target given in Ref.~\cite{Berlin:2018jbm}.

\subsection{Established limits and future prospects}

Before studying in detail the sensitivity of the \belletwo experiment to inelastic DM, let us briefly discuss existing constraints on the parameter space.
As mentioned above, electroweak precision observables constrain the kinetic mixing parameter $\epsilon$ irrespective of any couplings of the $A^\prime$ to dark sector states to be smaller than $\epsilon \lesssim 3 \times 10^{-2}$ for dark 
photon masses below the $Z$ mass~\cite{Hook:2010tw}.
In addition there are a variety of experimental probes which are sensitive to more specific signatures of the model that we consider, ranging from electron and proton beam dumps over low-energy colliders to direct detection experiments.

Fixed-target experiments are sensitive to inelastic DM via a number of different search strategies. In a recent analysis, NA64 obtained exclusion bounds on the production of DM particles through dark photon decays based on the resulting missing energy in the detector~\cite{NA64:2019imj}. The DM particles produced in the beam dump may also induce scattering in a far detector, leading to relevant bounds from LSND~\cite{deNiverville:2011it} and MiniBoonNE~\cite{Aguilar-Arevalo:2017mqx}. Finally, strong exclusion limits are obtained from the non-observation of $\chi_2$ decays in E137~\cite{Berlin:2018pwi}, NuCal and CHARM~\cite{Tsai:2019mtm}. All of these experiments are most sensitive to small dark photon masses, $m_{A'} \ll 1\,\mathrm{GeV}$, and are therefore complementary to the searches that we will discuss below. For the case of $\chi_2$ decays the reinterpretation of published bounds for different model parameters requires Monte Carlo simulations of the $\chi_2$ production and decays, which are beyond the scope of the present work. We will therefore only show these constraints for parameter combinations for which exclusion limits are readily available in the literature.

As we will discuss in more detail below, \bfactories such as \babar or \belletwo typically need an associated photon to be able to trigger on the production of a dark photon, $e^+ e^- \rightarrow \gamma A^\prime$, $A^\prime \rightarrow \chi_1 \chi_2$. 
If the $\chi_2$ state is sufficiently long-lived such that the decays of unstable $\chi_2$ happen outside of the detector, searches for a single photon in association with missing energy (so-called mono-photon searches) give relevant constraints on $A^\prime$ production. 
The reinterpretation of the BaBar mono-photon limit~\cite{Lees:2017lec} requires the evaluation of acceptances specific to that experiment, so we postpone the discussion of this limit to the next section.

In addition there are a large number of proposed future experiments which have projected sensitivities surpassing the current limits, see e.g.\ figure~7 of Ref.~\cite{Berlin:2018jbm} for a comprehensive compendium, including possible add-ons to the LHC such as FASER~\cite{Feng:2017uoz}, MATHUSLA~\cite{Chou:2016lxi}, and CODEX-b~\cite{Gligorov:2017nwh} or future beam dumps such as LDMX~\cite{Akesson:2018vlm} and SeaQuest~\cite{Berlin:2018pwi}. We will consider these projections more closely in section~\ref{SEC:discussion}.

\subsubsection*{Direct detection}

At tree-level, scattering of DM particles from the Galactic halo off a nucleon $N$ in direct detection experiments can only
proceed via the inelastic process $\chi_1 + N \to \chi_2 + N$. For
$\Delta \gtrsim 10^{-6} \, m_{\chi_1}$ the kinetic energy in the initial
state is insufficient to overcome the mass splitting between $\chi_1$
and $\chi_2$, such that inelastic scattering is forbidden. Nevertheless
the elastic scattering process $\chi_1 + N \to \chi_1 + N$ arises at the
one-loop level from diagrams with two dark photon exchanges. These
diagrams have recently been calculated in Ref.~\cite{Sanderson:2018lmj}
and we will briefly summarize the result here.

The box diagrams give a contribution to the Wilson coefficient $C_q$ of
the effective operator
\begin{equation}
  \mathcal{O}_q = \sum_q q_q^2 m_q \overline{q} q \overline{\chi_1} \chi_1 \; ,
\end{equation}
where $q_q$ and $m_q$ denote the electric charge and mass of the quarks.
One finds\footnote{We point out that Ref.~\cite{Sanderson:2018lmj}
assumes $M_{A^\prime} \gg m_t$ such that the mediator can be integrated out
before the top quark. As pointed out in a different context in
Refs.~\cite{Abe:2018emu,Ertas:2019dew}, this approach may give incorrect
results for smaller mediator masses. Nevertheless, a more accurate
estimate would require a two-loop calculation, which is well beyond the
scope of the present work. We will therefore use the results from
Ref.~\cite{Sanderson:2018lmj} for the estimates presented here.}
\begin{equation}
  C_q = \frac{4 \epsilon^2 e^2 \alpha_\text{D} m_{\chi_1}}{4 \pi m_{A^\prime}^4}
F_3\Big(\frac{m_{\chi_1}^2}{m_{A^\prime}^2}\Big)
\end{equation}
with
\begin{equation}
  F_3(x) = \frac{(8 x^2 - 4 x + 2) \log\left(\frac{\sqrt{1 - 4 x} + 1}{2
\sqrt{x}}\right) + \sqrt{1 - 4 x} (2 x + \log x)}{4  x^2 \sqrt{1 - 4 x}}
\; .
\end{equation}

At the hadronic scale the operator $\mathcal{O}_q$ matches onto the
DM--nucleon operator
\begin{equation}
  \mathcal{O}_N = \overline{N} N \overline{\chi_1} \chi_1
\end{equation}
with coefficient $C_N = 0.082 \, m_N \, C_q$. Note that this
result differs from the well-known formula $C_N \approx 0.3 \, m_N \,
C_q$ due to the extra factors of $q_q$ included in the definition of
$\mathcal{O}_q$. In terms of this coefficient, the spin-independent DM--nucleon
scattering cross section is simply given as
\begin{equation}
\sigma_N = \frac{4 \, \mu^2 C_N^2}{\pi} \; ,
\end{equation}
where $\mu$ is the reduced mass.

Since the loop-induced direct detection cross section is proportional to
$\alpha_\text{D}^2 \epsilon^4$, probing $\epsilon \ll 1$ is extremely
challenging. Furthermore, the sensitivity of direct detection
experiments is substantially suppressed for DM masses below a few
GeV.\footnote{At first sight, DM--electron scattering offers a promising
way to search for inelastic DM with sub-GeV masses. However, the
loop-induced DM--electron scattering cross section is suppressed relative
to the one for DM--proton scattering by a factor $m_e^4 / (m_N^2 \,
m_{\chi_1}^2)$, which renders DM--electron scattering irrelevant.} As a
result, we find that even future direct detection experiments like
SuperCDMS~\cite{Agnese:2016cpb} are not competitive with $e^+ e^-$
colliders. According to the official sensitivity projection of SuperCDMS, this experiment would be sensitive to $\alpha_\mathrm{D} \epsilon^2 \sim 10^{-3}$ for $m_\chi \sim 1\,\mathrm{GeV}$. Hence, for $\alpha_\mathrm{D} \leq 0.5$ SuperCDMS will not be able to improve upon the LEP bound $\epsilon < 3 \times 10^{-2}$. Hence we conclude that direct detection bounds are essentially irrelevant to our model, and therefore do not display them in our figures.

\section{Light inelastic dark matter at \texorpdfstring{\belletwo}{belletwo}}
\label{SEC:belletwo}

\begin{figure}[tb]
  \centering
  \includegraphics[width=0.7\textwidth]{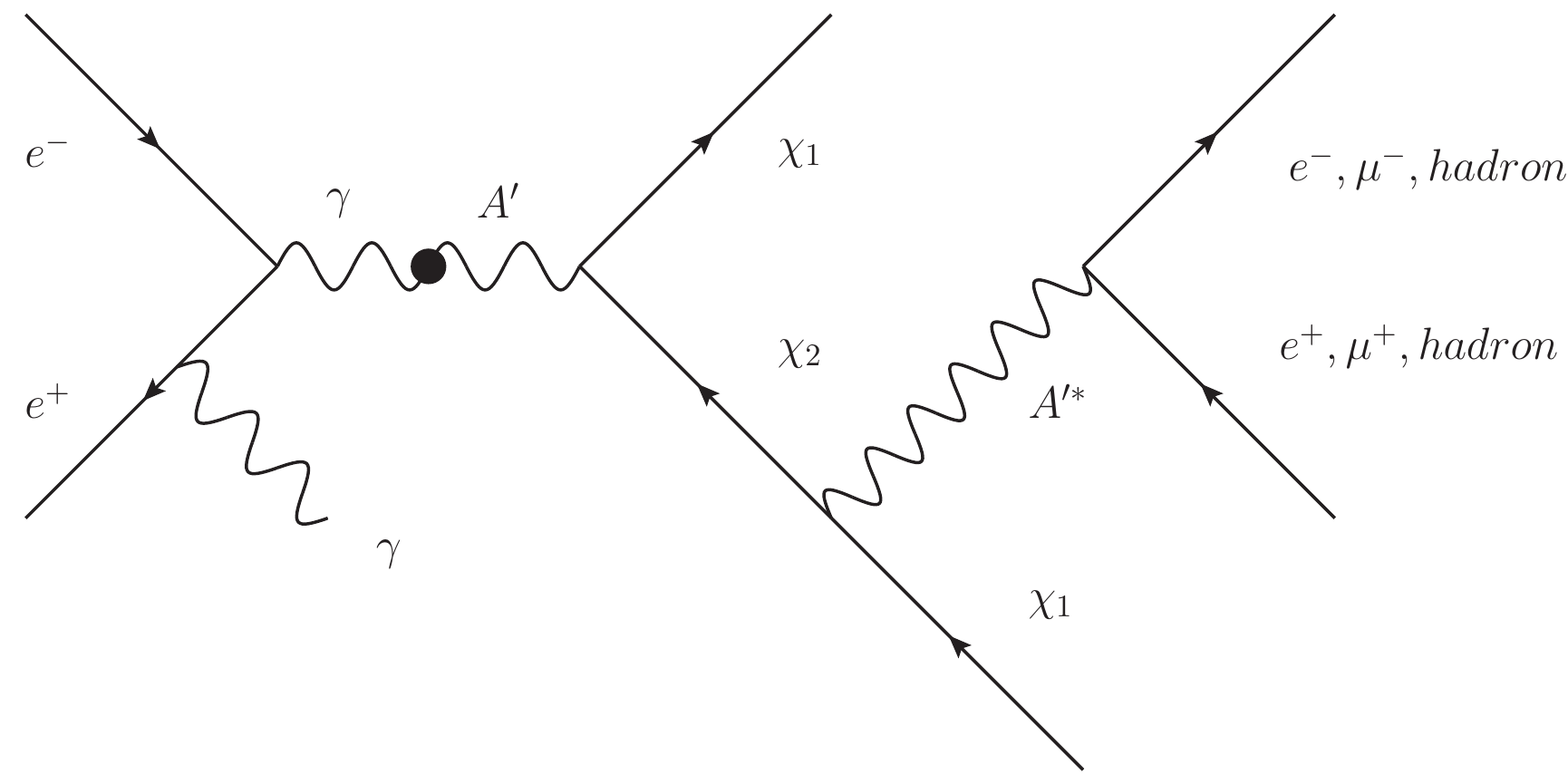}
  \caption{
    The Feynman diagram depicting the photon and displaced fermion signature in the context of the inelastic DM scenario.
    \label{fig:feynmandiagram}
  }
\end{figure} 

Broadly speaking, the inelastic DM model can produce two types of signatures in \belletwo, which both arise from the process shown in figure~\ref{fig:feynmandiagram}. If the $\chi_2$ produced via $e^+ e^- \to \gamma A^\prime (\to \chi_1 \chi_2)$ decays outside the detector, the final state is indistinguishable from the process $e^+ e^- \to \gamma A^\prime$, $A^\prime \to \text{invisible}$ usually searched for at $e^+ e^-$ colliders. The same signature arises if the $\chi_2$ decay vertex is inside the detector but the decay products have too low energy to be detected. If on the other hand the $\chi_2$ decay products are detected and the decay vertex can be reconstructed, one obtains a displaced signature. In this section we will first review the relevant aspects of the \belletwo experiment, present our implementation of the inelastic DM model and then discuss the sensitivity of \belletwo for both of these signatures.

\subsection{The \texorpdfstring{\belletwo}{belletwo} experiment}
The \belletwo experiment at the SuperKEKB accelerator is a second generation \bfactory and successor of the \belle and \babar experiments~\cite{Abe:2010gxa}. 
Construction was completed in early 2019. 
\superkekb is a circular asymmetric $e^+e^-$ collider with a nominal collision energy of $\sqrt{s} = \unit[10.58]{GeV}$. 
The design instantaneous luminosity is \hbox{$8\times10^{35}$\,cm$^{-2}$ s$^{-1}$}, which is about 40 times higher than at the predecessor collider KEKB.

The \belletwo detector is a large-solid-angle magnetic spectrometer. 
The following sub-detectors are particularly relevant for the searches described in this paper: 
a tracking system that consists of six layers of vertex detectors (VXD), including two inner layers of silicon pixel detectors (PXD)\footnote{During the first years of \belletwo only the first layer and a fraction of the second PXD layer are instrumented. We assume that this has a negligible effect for the searches described in this paper.} and four outer layers of silicon vertex detectors (SVD), and a 56-layer central drift chamber (CDC) which covers a polar angle region of $(17-150)^{\circ}$. 
The electromagnetic calorimeter (ECL) comprising CsI(Tl) crystals with an upgraded waveform sampling readout for beam background suppression covers a polar angle region of $(12-155)^{\circ}$ and is located inside a superconducting solenoid coil that provides a 1.5\,T magnetic field. 
The ECL has inefficient gaps between the endcaps and the barrel for polar angles between $(31.3-32.2)^{\circ}$ and $(128.7-130.7)^{\circ}$. 
An iron flux-return is located outside of the magnet coil and is instrumented with resistive plate chambers and plastic scintillators to mainly detect $K^0_L$ mesons, neutrons, and muons (KLM) that covers a polar angle region of $(25-145)^{\circ}$.

We study the \belletwo sensitivity for a dataset corresponding to an integrated luminosity of $\unit[20]{fb^{-1}}$ for consistency with~\cite{Kou:2018nap}. 
This dataset is expected to be recorded by \belletwo in early 2020.
To show the potential reach of \belletwo we also estimate the sensitivities for both the mono-photon signature and the displaced signature for the final dataset of $\unit[50]{ab^{-1}}$.
For the displaced signature we optimistically assume that the search remains background free even for very large luminosities.
For the mono-photon signature we scale the expected sensitivity $S(\epsilon)$ to the planned full integrated luminosity of $\unit[50]{ab^{-1}}$ using $S(\epsilon) \propto \sqrt[4]{\mathcal{L}}$.
This scaling is valid under the following assumptions: 
\begin{itemize}
\item The expected increase of beam induced background noise at highest luminosity and the resulting decrease in ECL energy resolution is negligible,
\item the expected increase in the number of background induced photons is not relevant for these searches as we assume that they can be rejected by timing and cluster-shape selections,
\item the triggers can be kept loose enough to achieve $\approx\unit[100]{\%}$ trigger efficiency,
\item the searches are dominated by statistical uncertainties. 
\end{itemize}

\subsection{Event generation}
We implemented the model into \texttt{FeynRules v2.3.32}~\cite{Alloul:2013bka} to generate a \texttt{UFO} model file~\cite{Degrande:2011ua}. To produce signal events we generate events for $e^+ e^- \to \chi_1 \chi_2 \gamma$ with \texttt{MadGraph5\textunderscore{}aMC@NLO v2.6.6}~\cite{Alwall:2014hca} and subsequently perform the decays $\chi_2 \to \chi_1 e^+ e^-$ and $\chi_2 \to \chi_1 \mu^+ \mu^-$ in \texttt{MadSpin}~\cite{Artoisenet:2012st}.
We use \texttt{MadGraph5\textunderscore{}aMC@NLO v2.6.6}~\cite{Alwall:2014hca} to calculate the total width (and hence the decay length) of $\chi_2$. We do not calculate decays to hadronic final states\footnote{In the mass range of interest to us, the majority of any hadronic final state will be composed of pions.} since hadrons will be treated as muons in all our analyses. Instead we rescale the partial decay width into muons by including the measured $R(s)$ values \cite{Tanabashi:2018oca}.

Since we always require $m_{A^\prime} > m_{\chi_1} + m_{\chi_2}$, the $A^\prime$ can never be on-shell in the $\chi_2$ decay and only three-body decays are allowed. 
Nevertheless the branching fractions are largely determined by the $A'$ branching ratios \cite{Ilten:2018crw}.
Note that the $\chi_2$ branching ratios are independent of $\alpha_\text{D}$ and $\epsilon$. For $m_{\chi_2} - m_{\chi_1} < 2 m_e$ the only kinematically allowed decays are $\chi_2 \to \chi_1 \overline{\nu} \nu$ and $\chi_2 \to \chi_1 \gamma \gamma \gamma$, which are highly suppressed.  
We conservatively assume that there is no significant contribution from charged hadronic two particle final states above $\Delta = \unit[1.2]{GeV}$. This is a good approximation since the dark photon branching ratio to $\pi^+ \pi^-$ or $K^+ K^-$ final states is subdominant above centre-of-mass energy/virtual dark photon mass of \unit[1.2]{GeV}~\cite{Liu:2014cma}.

We generate the events in the centre-of-mass frame with \hbox{$\sqrt{s} = \unit[10.58]{GeV}$}, then boost and rotate them to the \belletwo laboratory frame. The \belletwo beam parameters are \hbox{$E(e^+) = \unit[4.002]{GeV}$} and \hbox{$E(e^-) = \unit[7.004]{GeV}$} with a $\unit[41.5]{mrad}$ crossing angle between the beams and the z-axis.  In the laboratory frame the z-axis is along the bisector of the angle between the direction of the electron beam and the reverse direction of the positron beam. All cuts below refer to parameters in the lab frame unless noted otherwise. 

\begin{figure}[tb]
 \centering
 \includegraphics[width=.99\textwidth]{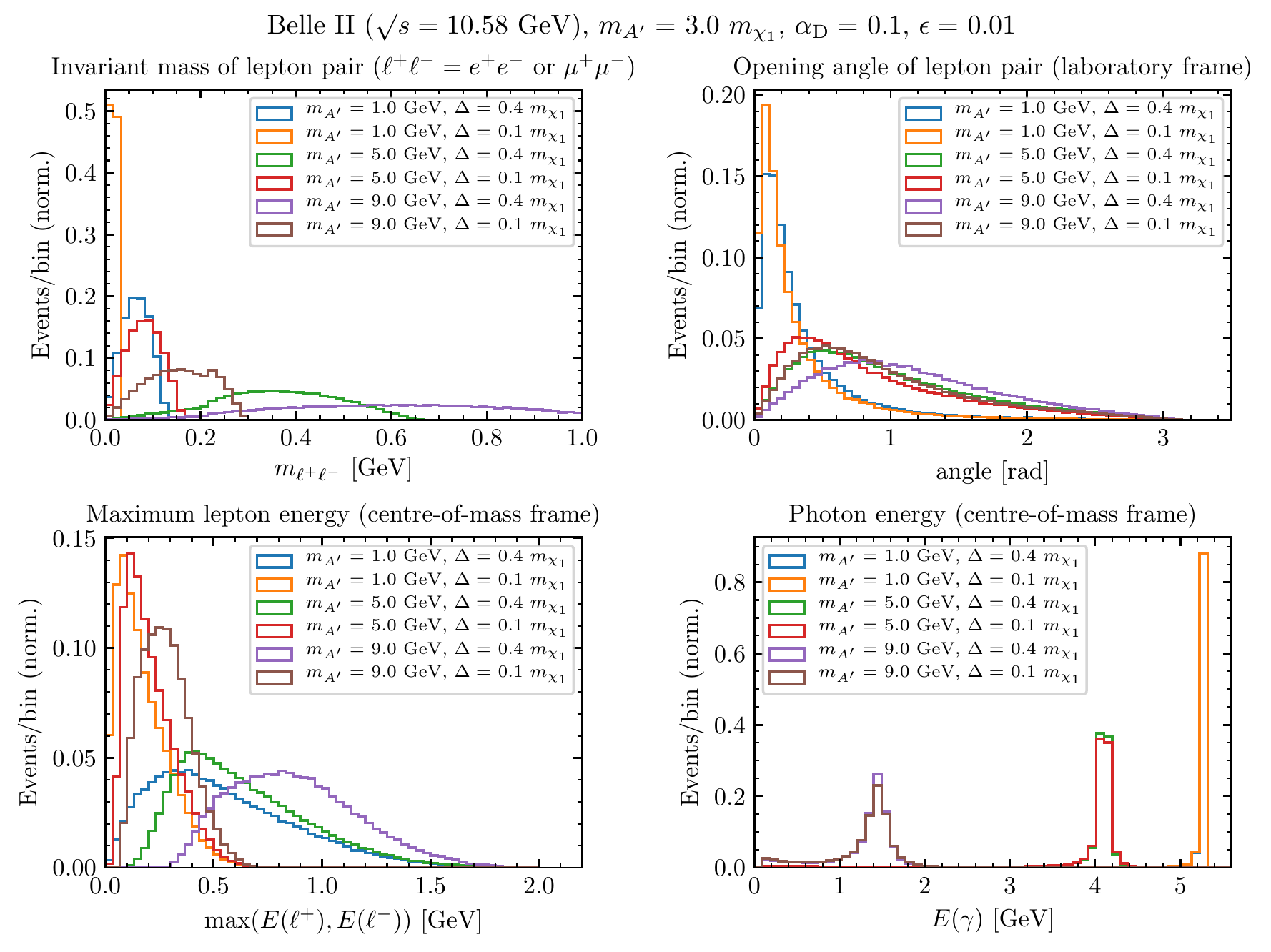}
 \caption{
 Histograms of various observables for our signal (top left: invariant mass of the lepton pair, top right: opening angle of the lepton pair, bottom left: maximum lepton energy, bottom right: photon energy). Note that the opening angle is given in the \belletwo lab frame, whereas the maximum lepton energy and the photon energy are given in the centre-of-mass frame in this figure and in the text. In the lower right panel, the curves for $\Delta = 0.4\, m_{\chi_1}$ and $\Delta = 0.1\, m_{\chi_1}$ for $m_{A^\prime} = \unit[1]{GeV}$ completely overlap.}\label{fig:histograms}
\end{figure}

A collection of interesting observables are shown in figure~\ref{fig:histograms}. All plots are at the generator level, with no detector smearing applied. For signal generation we apply a cut of \hbox{$E_\text{CMS}(\gamma) > \unit[0.1]{GeV}$} and a maximal rapidity of the photon \hbox{$\eta_\text{max} = 2.028698$} in the centre-of-mass frame. 

We point out a number of relevant features:
\begin{itemize}
\item The invariant mass of the di-lepton pair must satisfy the requirement $m_{\ell^+ \ell^-} \leq \Delta$ and typically peaks at around half of this value.
 \item The opening angle of the di-lepton pair in the laboratory frame depends sensitively on the boost (and hence the mass) of $\chi_2$, i.e.\ lighter $\chi_2$ will have higher boost and hence lead to smaller opening angles of the di-lepton pair.
 \item The maximum lepton energy is a combination of the two previous effects, i.e.\ it increases both with the mass splitting and with the boost of the $\chi_2$.
 \item The photon in the centre-of-mass frame is essentially mono-energetic (with some broadening due to the finite width of the dark photon).
\end{itemize}

\subsection{Mono-photon signature}

To rescale the expected \belletwo mono-photon sensitivity for a $\unit[20]{fb^{-1}}$ data set~\cite{Kou:2018nap}, we calculate the acceptances on our signal sample as follows.
We assume that events satisfy the mono-photon selection criteria if $E_\text{CMS}(\gamma) > \unit[2.0]{GeV}$, and apply a polar angle selection that depends on the photon energy and the dark photon mass $m_{A^\prime}$~\cite{Kou:2018nap}. For $m_{A^\prime} < \unit[6.0]{GeV}$ we select events if $\theta_{\text{min}}^{\text{low}} < \theta(\gamma) < \theta_{\text{max}}^{\text{low}}$ with 
\begin{align}
\theta_{\text{min}}^{\text{low}} & = 5.399^{\circ} E_\text{CMS}(\gamma)^2/\text{GeV}^{2} - 58.82^{\circ}E_\text{CMS}(\gamma)/\text{GeV} + 195.71^{\circ},\\
\theta_{\text{max}}^{\text{low}}& = -7.982^{\circ}E_\text{CMS}(\gamma)^2/\text{GeV}^{2} + 87.77^{\circ}E_\text{CMS}(\gamma)/\text{GeV} - 120.6^{\circ}.
\end{align}
For $m_{A^\prime} \geq \unit[6.0]{GeV}$ we select events if $\theta_{\text{min}}^{\text{high}} < \theta(\gamma) < \theta_{\text{max}}^{\text{high}}$ with 
\begin{align}
\theta_{\text{min}}^{\text{high}} & = 3.3133^{\circ} E_\text{CMS}(\gamma)^2/\text{GeV}^{2} -33.58^{\circ}E_\text{CMS}(\gamma)/\text{GeV} + 108.79^{\circ},\\
\theta_{\text{max}}^{\text{high}}& = -5.9133^{\circ}E_\text{CMS}(\gamma)^2/\text{GeV}^{2} +54.119^{\circ}E_\text{CMS}(\gamma)/\text{GeV} + 13.781^{\circ}.
\end{align}

We consider electrons, muons, and hadrons as possible decay products of the $\chi_2$ and reject the event if the $\chi_2$ decay satisfies at least one of the veto criteria outlined in Table~\ref{tab:BelleIIvetoes}. 
For the veto criteria that we impose on the $\chi_2$ decay vertex we distinguish between the calorimeter/drift chamber and the muon system. 
The muon system criteria include a small region of phase space where the $\chi_2$ decays between the calorimeter and muon system in the forward or backward directions. In practice a decay into electrons would not be detected in this region, but this effect is negligible for the sensitivity estimation.

We then calculate the expected sensitivity in terms of the kinetic mixing $\epsilon$ in the inelastic DM model (see section~\ref{SEC:theory_model}), $\epsilon_\text{exp}^\text{iDM}$, based on the expected mono-photon sensitivity from Ref.~\cite{Kou:2018nap} (figure~209), $\epsilon_\text{exp}^{\text{mono-}\gamma}$, via the following rescaling:
\begin{equation}
\epsilon_\text{exp}^\text{iDM} = \epsilon_\text{exp}^{\text{mono-}\gamma} \sqrt{\frac{\text{number of events selected based on photon criteria only}}{\text{number of events using all selection criteria}}} .
\end{equation}

\begin{table}[tb]
 \centering
 \caption{Vetoes on electrons, muons, and hadrons used for the \belletwo mono-photon analysis rescaling. 
 The variables $\theta_\text{lab}$, $z$, and $R_{xy}$ refer to the $\chi_2$ decay vertex in the laboratory frame.
$\theta_\text{lab}$ is the polar angle between the decay vertex and positive $z$ direction, and $R_{xy}$ is the distance between the $z$ axis and the decay vertex in the plane perpendicular to the $z$ axis.  \label{tab:BelleIIvetoes}}
\begin{tabular}{lll}
 particle type & calorimeter/drift chamber & muon system  \\ 
 \hline \hline 
 \multirow{4}{4em}{electrons} & (i) either $E(e^-)$ or $E(e^+)$  $> \unit[150]{MeV}$  & (i)  $E(e^+) + E(e^-) > \unit[300]{MeV}$ \\
                         & (ii) and $17^\circ < \theta_\text{lab} < 150^\circ$   & (ii) and $25^\circ < \theta_\text{lab} < 145^\circ$   \\
                         & (iii) and $\unit[-112]{cm} < z < \unit[206]{cm}$      & (iii) and $\unit[-300]{cm} < z < \unit[400]{cm}$      \\
                         & (iv) and $R_{xy} < \unit[135]{cm}$                    & (iv) and $R_{xy} < \unit[300]{cm}$                    \\
 \hline                        
 \multirow{2}{4em}{muons} & (i) either $p(\mu^-)$ or $p(\mu^+) > \unit[150]{MeV}$  & (i) $p(\mu^+) + p(\mu^-) > \unit[300]{MeV}$ \\
                     & (ii)--(iv) as for electrons               & (ii)--(iv) as for electrons \\
 \hline
 hadrons & treat as muons & treat as muons \\
 \hline \hline\
\end{tabular}
\end{table}

Along the same lines we can perform a reinterpretation of the \babar mono-photon analysis~\cite{Lees:2017lec} for the inelastic DM model, in order to compare the sensitivity of \belletwo with existing constraints.
This analysis, however, does not use cuts on simple quantities but a multivariate analysis, so a straight-forward reinterpretation is not possible. 
To model the analysis as closely as possible, we use the same procedure as described above,\footnote{The \babar beam parameters are $E(e^+) = \unit[3]{GeV}$ and $E(e^-) = \unit[9.5]{GeV}$ with no crossing angle between the beams, and the forward $z$ direction pointing in the negative $e^+$ direction.} but we use the vetoes described in Table~\ref{tab:BaBarvetoes} to specify whether an event will be rejected by the analysis and to calculate the mono-photon acceptance. The photon selection criteria we use are $E_\text{CMS}(\gamma) > \unit[2.0]{GeV}$ and $32.5^\circ < \theta(\gamma) < 99^\circ$.

\begin{table}[bt]
 \centering
 \caption{Vetoes on electrons, muons, and hadrons used for the \babar mono-photon analysis. The variables $\theta_\text{lab}$, $z$, and $R_{xy}$ refer to the $\chi_2$ decay vertex in the laboratory frame, same definitions as in Table~\ref{tab:BelleIIvetoes}. \label{tab:BaBarvetoes}}
\begin{tabular}{lll}
 particle type & calorimeter/drift chamber & muon system  \\ 
 \hline \hline 
 \multirow{4}{4em}{electrons} & (i) either $E(e^-)$ or $E(e^+)$  $> \unit[150]{MeV}$  & (i)  $E(e^+) + E(e^-) > \unit[300]{MeV}$ \\
                         & (ii) and $17^\circ < \theta_\text{lab} < 142^\circ$   & (ii) and $20^\circ < \theta_\text{lab} < 150^\circ$   \\
                         & (iii) and $\unit[-113]{cm} < z < \unit[185]{cm}$      & (iii) and $\unit[-223]{cm} < z < \unit[297]{cm}$      \\
                         & (iv) and $R_{xy} < \unit[102]{cm}$                    & (iv) and $R_{xy} < \unit[243]{cm}$                    \\
 \hline                        
 \multirow{2}{4em}{muons} & (i) either $p(\mu^-)$ or $p(\mu^+) > \unit[150]{MeV}$  & (i) $p(\mu^+) + p(\mu^-) > \unit[300]{MeV}$ \\
                     & (ii)--(iv) as for electrons               & (ii)--(iv) as for electrons \\
 \hline
 hadrons & treat as muons & treat as muons \\
 \hline \hline\
\end{tabular}
\end{table}

As for the \belletwo case, the muon system criteria includes a small region of phase space where the $\chi_2$ decays between the calorimeter and muon system in the forward directions. In practice, a decay to electrons would not be detected in this region, but this effect is negligible for the sensitivity estimation. 

As we will see below, even with an integrated luminosity of $\mathcal{L}_{\text{int}}=20\,\text{fb}^{-1}$ \belletwo can improve substantially on the existing constraint from \babar (integrated luminosity $\mathcal{L}_{\text{int}}=53\,\text{fb}^{-1}$) at lower masses $m_{\chi_1}$ because of a more hermetic calorimeter in \belletwo.

\subsection{Displaced signature}

\begin{figure}[tb]
 \centering
 \includegraphics[angle=90,height=8cm]{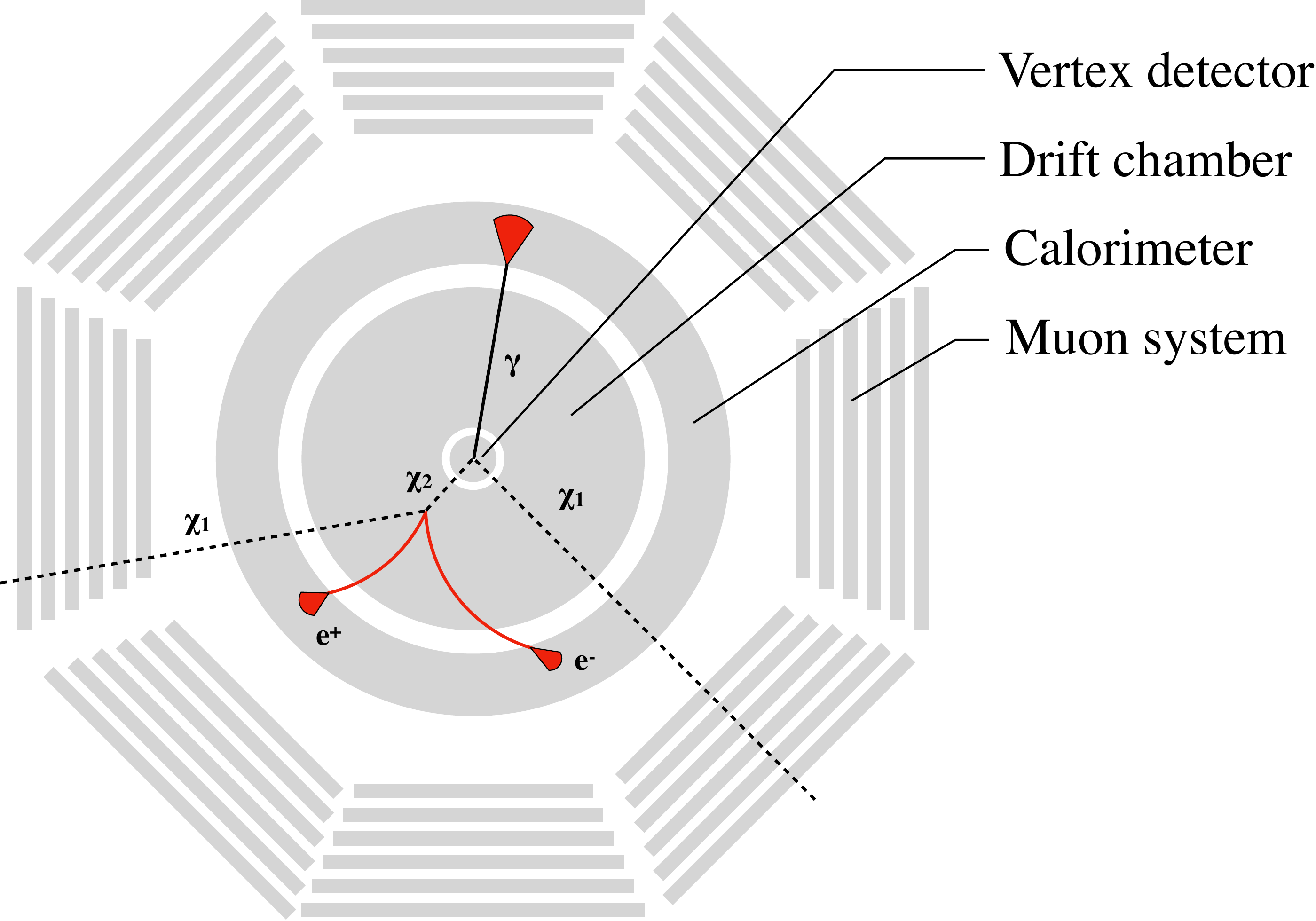}
 \caption{Schematic view of the \belletwo detector ($xy$-plane) and example displaced signature.}\label{fig:schematicidm}
\end{figure}

Let us now take a closer look at the characteristic signature of inelastic DM: a displaced lepton or hadron pair (pions and kaons) in association with a single photon. An illustrative example of what this signature may look like in the \belletwo detector is shown in figure~\ref{fig:schematicidm}.
In the following we will take a closer look at this signature, identify possible backgrounds and event selection criteria, and discuss the challenge of triggers.

\subsubsection{Backgrounds}

SM processes can produce a (displaced) lepton or meson pair, a hard photon, and missing energy only if particles are out of the detector acceptance or if secondary processes contribute.
We consider the following backgrounds, where non-reconstructed particles are given in parentheses:
\begin{enumerate}
\item \textit{Direct} radiative lepton and meson pair production with two additional initial or final state radiation photons ($e^+e^-\to e^+e^-\gamma (\gamma)$, $e^+e^-\to \mu^+\mu^-\gamma (\gamma)$, $e^+e^-\to \pi^+\pi^-\gamma (\gamma)$), where one of the photons is out of the detector acceptance,
\item \textit{photon conversion} $\gamma \to e^+e^-$  from direct radiative electron pair production ($e^+e^-\to (e^+e^-)\gamma \gamma$) where both primary electrons are out of detector acceptance, or from radiative photon pair production ($e^+e^-\to \gamma \gamma (\gamma)$) where one photon is out of acceptance,
\item \textit{meson decays}, e.g. $e^+e^-\to \phi \gamma, \phi \to K^0_S (K^0_L), K^0_S\to \pi^+\pi^-$.
\end{enumerate}

Since the cross section of radiative Bhabha scattering is orders of magnitude larger than muon- or pion-pair production, we assume that $e^+e^-\to e^+e^-\gamma$ and $e^+e^-\to \gamma \gamma \gamma$ are the dominant backgrounds. 

Unlike searches in high multiplicity hadronic events \cite{Lees:2015rxq}, we expect negligible background from wrong track combinations that could fake displaced signatures.

\subsubsection{Event selection}

The strongest background rejection can be achieved by requiring a displaced vertex. We assume that the \belletwo detector can be split into five different regions in the azimuthal  $xy$-plane for the lepton pair vertex location, where $R_{xy}$ is the distance between the $z$ axis and the decay vertex in the plane perpendicular to the $z$ axis. 
\begin{enumerate}
\item $0\,\text{cm}\leq R_{xy} \leq 0.2\,\text{cm}$: The vertex location is very close to the nominal interaction point. We expect prohibitively large  prompt SM backgrounds.
\item $0.2\,\text{cm} < R_{xy} \leq 0.9\,\text{cm}$: The vertex location is inside the beam pipe, but outside of the interaction region. We expect excellent vertex reconstruction efficiency and negligible SM backgrounds.
\item $0.9\,\text{cm} < R_{xy} \leq 17\,\text{cm}$: The vertex location is inside the region covered by the VXD. We expect very good vertex reconstruction efficiency, but a sizeable background from photon conversions due to the material in this detector region. The estimation of the background is beyond the scope of this paper. We expect that selections based on the invariant mass of the lepton pair, or opening angle requirements of the two leptons could reduce the background significantly and this region could be included in a future analysis also for electron/positron final states. 
\item $17\,\text{cm} < R_{xy} \leq  60\,\text{cm}$: The vertex location is outside the VXD but inside the CDC. We expect that loose selections on the invariant mass of the lepton pair, or opening angle requirements of the two leptons, can reduce the background from photon conversion to a negligible level. 
\item $60\,\text{cm} < R_{xy} $: The vertex location is inside the CDC but the tracking efficiency is too low, or the vertex location is outside of any tracking detector acceptance.
\end{enumerate}

In the following we assume that \textit{direct} background can be completely rejected by removing events from region 1). 
\textit{Conversion background} can be reduced significantly by avoiding regions with high material density and requiring good vertex reconstruction efficiency, which we achieve by restricting the analysis to regions 2) and 4). 
To further reduce photon conversion backgrounds we require the invariant lepton pair mass $m_{\ell\ell} \geq \unit[0.03]{GeV}$, and an opening angle of at least 0.1\,rad between the leptons. 
We conservatively assume the efficiencies given in Table~\ref{tab:regionsR}. 
A more realistic analysis using the full \belletwo reconstruction information and a detailed material model, may allow to extend the analysis to include region 3) also for electron/positron final states in the future.

We veto the invariant mass region around the $K^0_S$ mass to reject \emph{meson decay} backgrounds for the muon and hadron final state. 
We assume that the meson decay background from two misidentified pions is negligible and do not include this background for the electron final state. 
We furthermore require a hard photon with $E_\text{lab}(\gamma) > \unit[0.5]{GeV}$ in the CDC acceptance of the detector.\footnote{The \belletwo calorimeter covers a slightly large polar angle range, but the material density outside of the tracking detector is too high for a good energy resolution.} All selections used in the analysis are summarized in Table~\ref{tab:cuts}. Note that for all detector regions we require minimal (transverse) momenta of all charged particles to ensure maximal tracking efficiency. We do not use missing energy information in this work but note that this could be used by a future analysis to further reduce backgrounds from SM processes.

A final selection requirement concerns the energy of the visible photon in the centre-of-mass frame. If the assumed dark photon mass $m_{A^\prime}$ is smaller than about $10 \, \mathrm{GeV}$, the dominant contribution stems from events where the dark photon is produced on-shell. To select these events and suppress background we require the energy of the visible photon in the centre-of-mass frame to lie within $\pm 10\%$ of the value dictated by energy conservation: $E_0 = (s-m_{A^\prime}^2 ) / (2 \sqrt{s})$. For larger dark photon masses, on the other hand, off-shell production becomes increasingly important and the photon energy will not exhibit a peak but instead rise steadily towards the low-energy threshold on $E_\text{lab}(\gamma)$ imposed above. This continuous photon spectrum makes both background rejection and the identification of a positive signal more difficult. Nevertheless, we can make a simple estimate of the sensitivity of \belletwo in the off-shell region by applying exactly the same selection requirement as for $m_{A^\prime} = 10\,\mathrm{GeV}$ (i.e.\ $0.507 \,\mathrm{GeV} < E_\text{CMS}(\gamma) < 0.620\,\mathrm{GeV}$), for which backgrounds can be assumed to be negligible. It is conceivable that a weaker requirement on $E_\text{CMS}(\gamma)$ would be sufficient to remove background while increasing the signal acceptance, but a detailed optimisation of the event selection for the off-shell region is beyond the scope of this work.

\begin{table}[bt]
\centering
 \caption{Regions and corresponding detection efficiencies used in the displaced analysis. $R_{xy}$ is the distance between the $z$ axis and the decay vertex in the plane perpendicular to the $z$ axis.
 \label{tab:regionsR}}
 \begin{tabular}{lcc}
  \multirow{2}{6em}{particle type} & low-$R_{xy}$ region & high-$R_{xy}$ region \\
                                    & (100\% detection eff.) & (30\% detection eff.) \\
  \hline \hline
  electrons & $\unit[0.2]{cm} < R_{xy} \leq \unit[0.9]{cm}$ & $\unit[17.0]{cm} < R_{xy} \leq \unit[60.0]{cm}$ \\
  muons     & $\unit[0.2]{cm} < R_{xy} \leq \unit[17.0]{cm}$ & $\unit[17.0]{cm} < R_{xy} \leq \unit[60.0]{cm}$ \\
 \end{tabular}

\end{table}

\begin{table}[bt]
\centering
 \caption{Selections used in the displaced vertex analysis. 
 \label{tab:cuts}}
 \begin{tabular}{ll}
  cut on  & value \\
  \hline \hline
   \multirow{2}{8em}{decay vertex} & (i) $\unit[-55]{cm} \leq z \leq \unit[140]{cm}$\\ 
                                         & (ii) $ 17^\circ\leq \theta_\text{lab} \leq 150^\circ$ \\
   \hline
   \multirow{3}{8em}{electrons} &  (i) both $p(e^+)$ and $p(e^-) > \unit[0.1]{GeV}$\\
             &  (ii) opening angle of pair $> 0.1$ \,rad\\
             &  (iii) invariant mass of pair $m_{ee}> \unit[0.03]{GeV}$\\
             \hline
   \multirow{4}{8em}{muons}     &  (i) both $p_\text{T}(\mu^+)$ and $p_\text{T}(\mu^-) > \unit[0.05]{GeV}$\\ 
             &  (ii) opening angle of pair $> 0.1$ \,rad\\
             &  (iii) invariant mass of pair $m_{ll} > \unit[0.03]{GeV}$\\
             &  (iv)  $m_{ll} < \unit[0.480]{GeV}$ or $m_{ll} > \unit[0.520]{GeV}$ \\
             \hline
   \multirow{3}{8em}{photons} & (i) $E_\text{lab} > \unit[0.5]{GeV}$ \\ 
                              & (ii) $ 17^\circ\leq \theta_\text{lab} \leq 150^\circ$  \\
                              & (iii) $0.9 \, E_0 \leq E_\text{CMS} \leq 1.1 \, E_0$, where $E_0 = (s-m_{A^\prime}^2 ) / (2 \sqrt{s})$ \\
   \hline \hline
 \end{tabular}
\end{table}

\subsubsection{Triggers}

So far we have assumed that all interesting events will be recorded by \belletwo and can be used for further analysis.
The displaced signature is however difficult to trigger, since it produces only a small number of final state particles.
We investigate three different triggers that are currently available at \belletwo and we study an additional displaced vertex trigger that is not available yet.

\paragraph{2\,GeV energy}
A calorimeter-only based trigger which requires at least one calorimeter cluster of $E_\text{CMS}(\gamma) > 2$\,GeV and $20^\circ < \theta_\text{lab}  < 139^\circ$ will be efficient for low mass dark photons $A^\prime$ for any mass splitting.
This trigger would also work for electrons of sufficiently high energy, but even for heavy dark photons and large mass splittings, the electrons rarely exceed $E_\text{CMS} > 2$\,GeV and will not pass this trigger. 

\paragraph{Three isolated clusters}
A calorimeter-only based trigger which requires at least three isolated calorimeter clusters will be efficient if both the photon and the two charged particles deposit enough energy in the calorimeter. 
The trigger requires at least one cluster with $E > 0.3$\,GeV, two other clusters with $E > 0.18$\,GeV, and $20^\circ < \theta_\text{lab}  < 139^\circ$, isolated by at least 30\,cm.
We assume that this trigger is available for low luminosity running of \belletwo only and that it will be prescaled by a factor 10 for the full dataset.

\paragraph{Two track triggers}
For events in signal regions 1), 2), or 3) the nominal two track trigger will be efficient if the transverse momentum $p_\text{T}>500$\,MeV for both tracks and if the azimuthal opening angle in the lab system is larger than $\Delta \phi > 90\,^\circ$. We assume that the two track trigger is not efficient in region 4).

\paragraph{Displaced vertex}
A trigger that is sensitive for displaced vertices will give the best sensitivity up to highest possible masses of $m_{\chi_1}$.
We assume that similar sensitivities as for the offline vertex reconstruction can be achieved in regions 3) and 4), that such a trigger would not require the additional presence of calorimeter activity, and that the displaced vertex trigger is not efficient in regions 1) and 2).

\subsection{Expected sensitivity}

\begin{figure}[tb]
  \centering
  \includegraphics[width=0.99\textwidth]{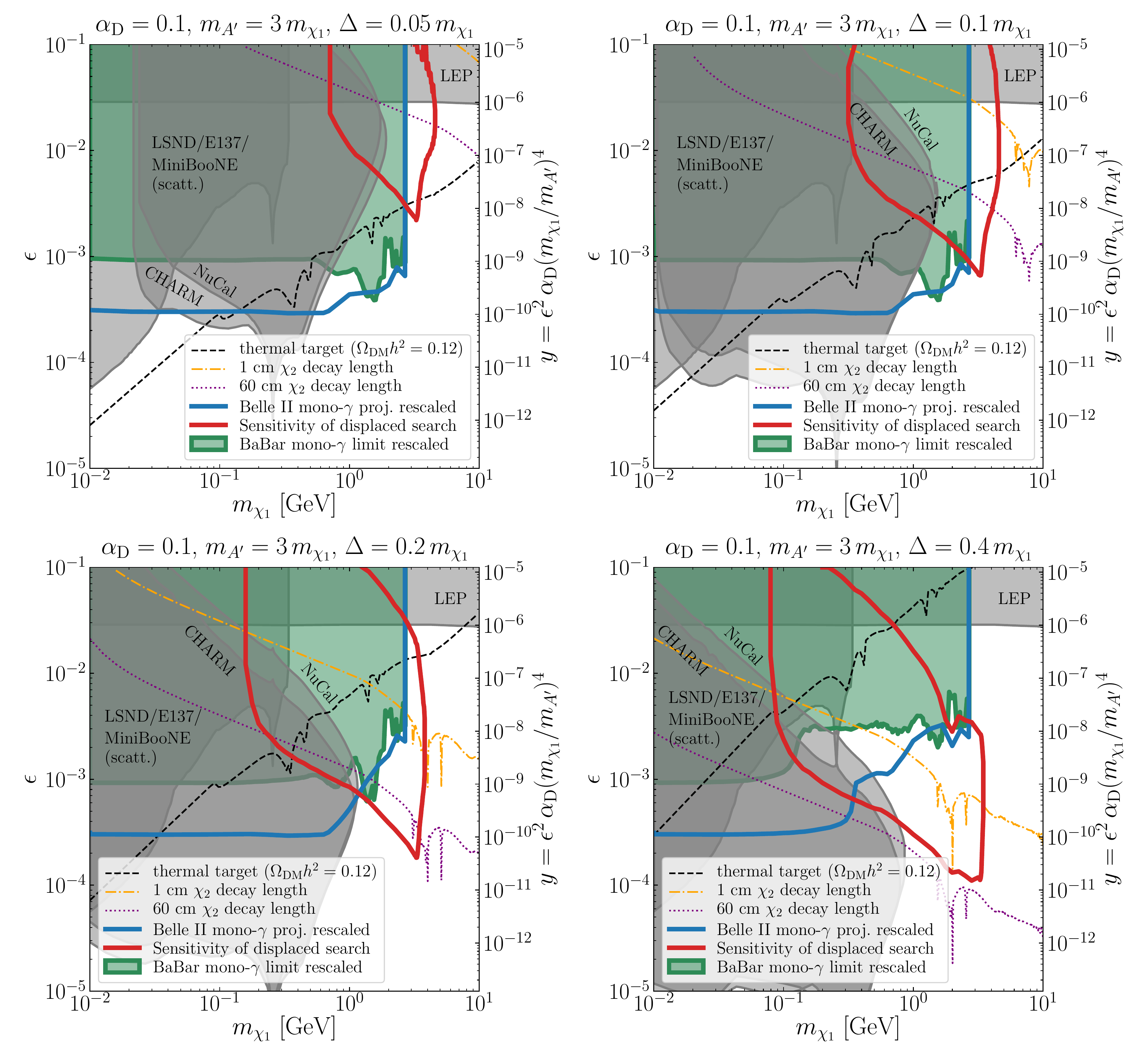}
  \caption{
    Sensitivity of \belletwo to the parameter space of inelastic DM for an integrated luminosity of 20 fb$^{-1}$ for $m_{A^\prime} = 3\, m_{\chi_1}$. 
    \label{fig:sensitivity3p0}
  }
\end{figure} 

\begin{figure}[tb]
  \centering
  \includegraphics[width=0.99\textwidth]{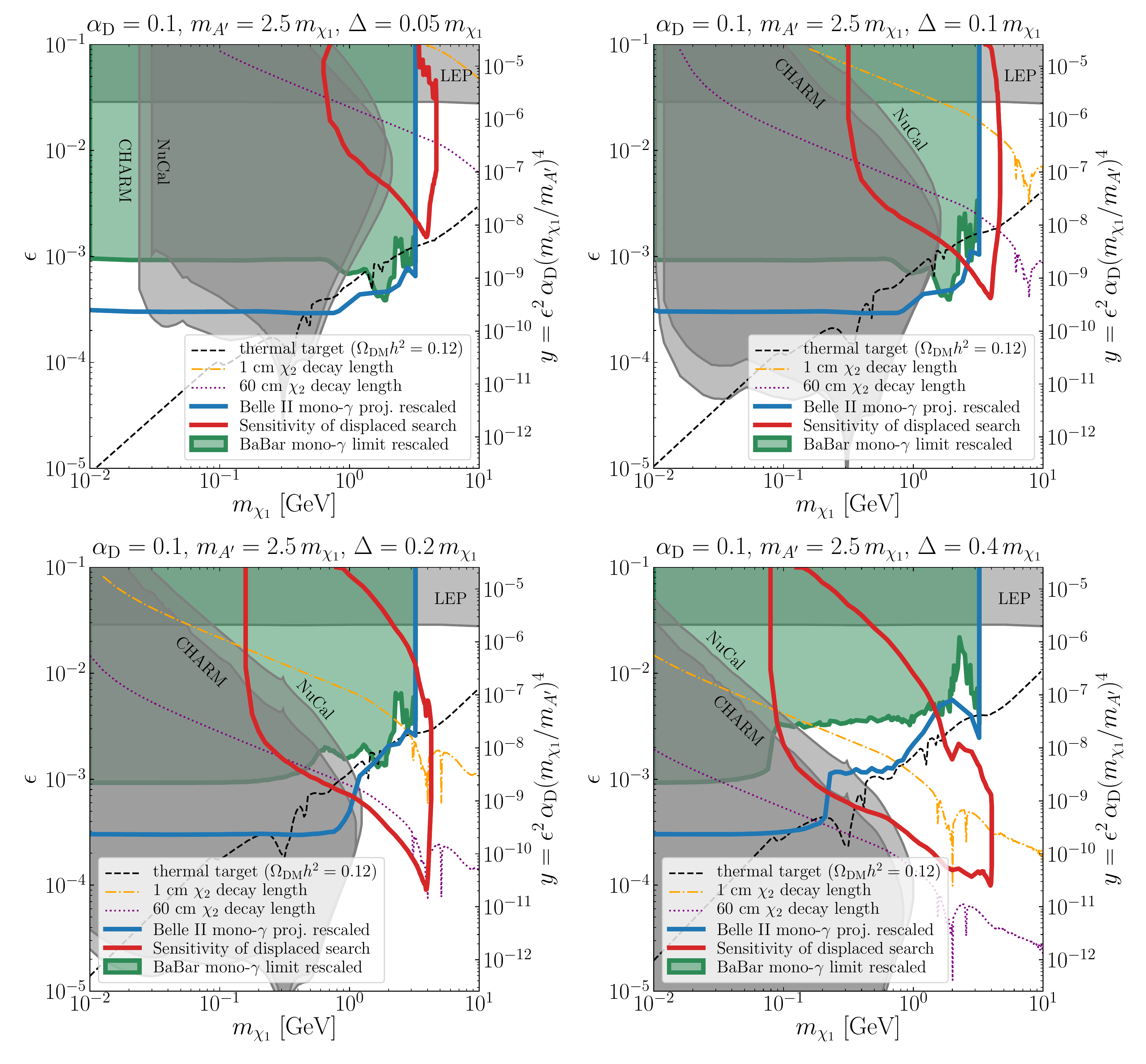}
  \caption{
    Sensitivity of \belletwo to the parameter space of inelastic DM for an integrated luminosity of 20 fb$^{-1}$  for $m_{A^\prime} = 2.5\, m_{\chi_1}$. 
    \label{fig:sensitivity2p5}
  }
\end{figure} 

Our main results are summarized in Figs.~\ref{fig:sensitivity3p0} and \ref{fig:sensitivity2p5}, which compare the sensitivity of our proposed search for displaced decays of inelastic DM with existing constraints as well as with the sensitivity of the mono-photon search. 
In both figures we vary the DM mass $m_{\chi_1}$ and the kinetic mixing parameter $\epsilon$ explicitly, while fixing the ratio of the dark photon mass and DM mass to $m_{A^\prime} = 3\, m_{\chi_1}$ and  $m_{A^\prime} = 2.5\, m_{\chi_1}$, respectively. The different panels in each figure correspond to different values of $\Delta$. We furthermore indicate the thermal target (black dashed) and examples of parameter combinations corresponding to fixed proper decay length of $\chi_2$. We show 90~\% C.L.\ contours which correspond to an upper limit of 2.3 events in the case of no background. Our chosen confidence level allows us to readily compare the sensitivity of our displaced search to the mono-photon limit from BaBar and the mono-photon sensitivity for \belletwo. 
Note that LSND/E137/MiniBooNE limits are available only for $m_{A^\prime} = 3\, m_{\chi_1}$.

We make the following observations: For small mass splitting $\Delta$, corresponding to large decay length of $\chi_2$ the bound from \babar and the projected sensitivity of the mono-photon search at \belletwo are very similar to the ones obtained for invisibly decaying dark photons, because the $\chi_2$ simply escapes from the detector before decaying. As soon as the decay length of the $\chi_2$ becomes comparable to the size of the detector, the sensitivity of these searches is significantly suppressed. Note that the bound does however not disappear entirely even for very short-lived $\chi_2$. The reason is that there always is a non-zero probability that the particles produced in the $\chi_2$ decay have very little transverse momentum (i.e.\ they travel in the direction of the beam pipe) and will not be reconstructed, so that the event resembles a single-photon event. The \babar bound that we obtain is therefore considerably stronger than the one from Ref.~\cite{Mohlabeng:2019vrz}, where no requirement on the angle $\theta_\text{lab}$ of the vertex is imposed.

As expected, the search for displaced decays performs best precisely in the region of parameter space where the mono-photon signal is suppressed and promises substantial improvements in particular for large mass splitting $\Delta$. But even for small mass splitting there is substantial room for improvement at large DM masses, corresponding to photon energies that would be too small to be observed in the absence of an additional lepton pair. Indeed, the sensitivity of the search for displaced decays extends even into the off-shell region, where $m_{A'} > \sqrt{s}$. In this region the energy of the visible photon is no longer mono-energetic and peaks at $E(\gamma) \to 0$, making the conventional strategy to perform a bump hunt to search for dark photons impossible. In this region the presence of a displaced lepton pair is therefore essential.

\begin{figure}[tb]
  \centering
  \includegraphics[width=0.99\textwidth]{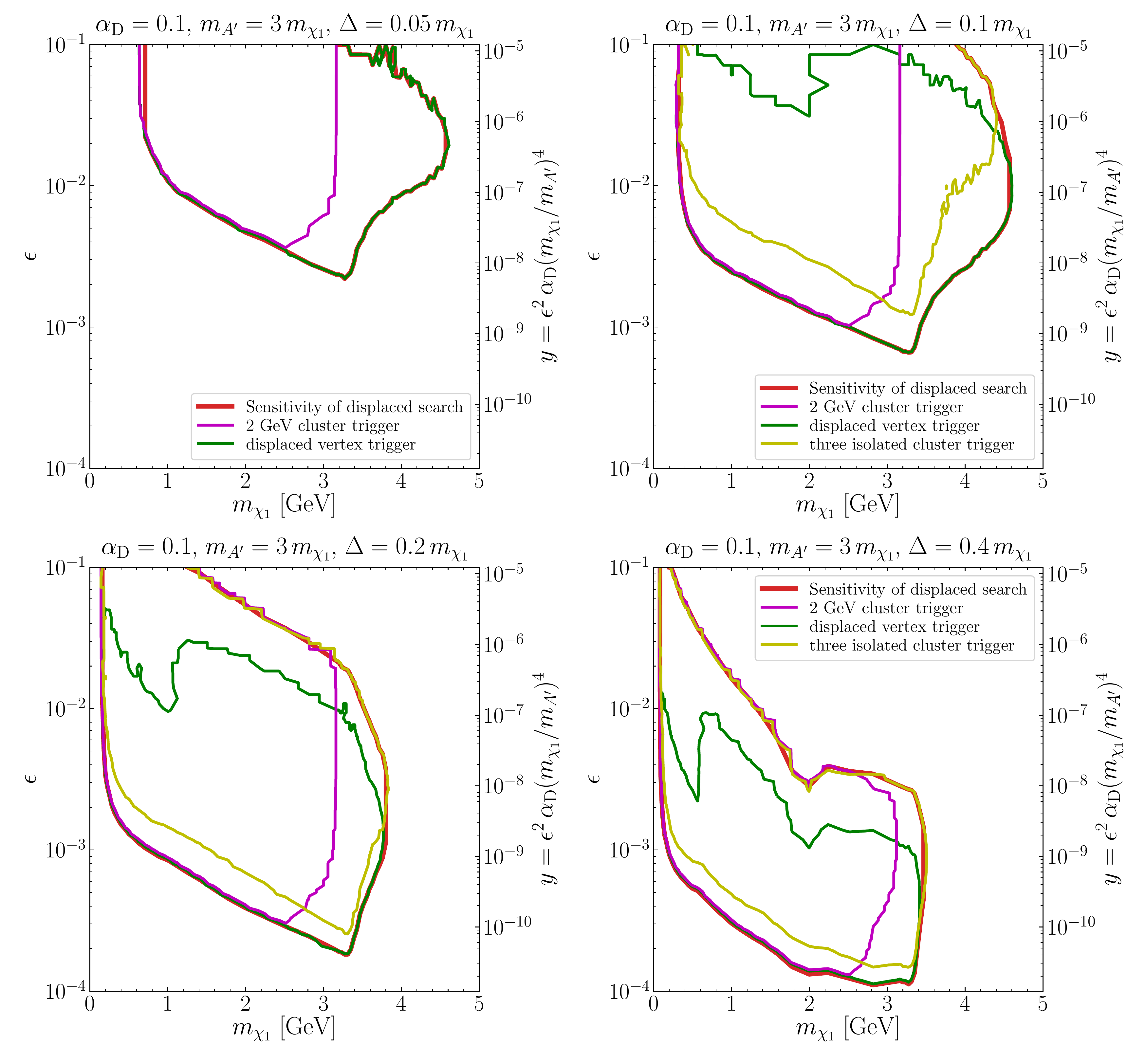}
  \caption{
    Sensitivity of the displaced search (same as figure~\ref{fig:sensitivity3p0}, but with linear horizontal axis), overlaid with the regions where the various triggers are efficient for an integrated luminosity of 20 fb$^{-1}$  for $m_{A^\prime} = 3.0\, m_{\chi_1}$.
    \label{fig:sensitivity3p0_trigger}
  }
\end{figure} 

\begin{figure}[tb]
  \centering
  \includegraphics[width=0.5\textwidth]{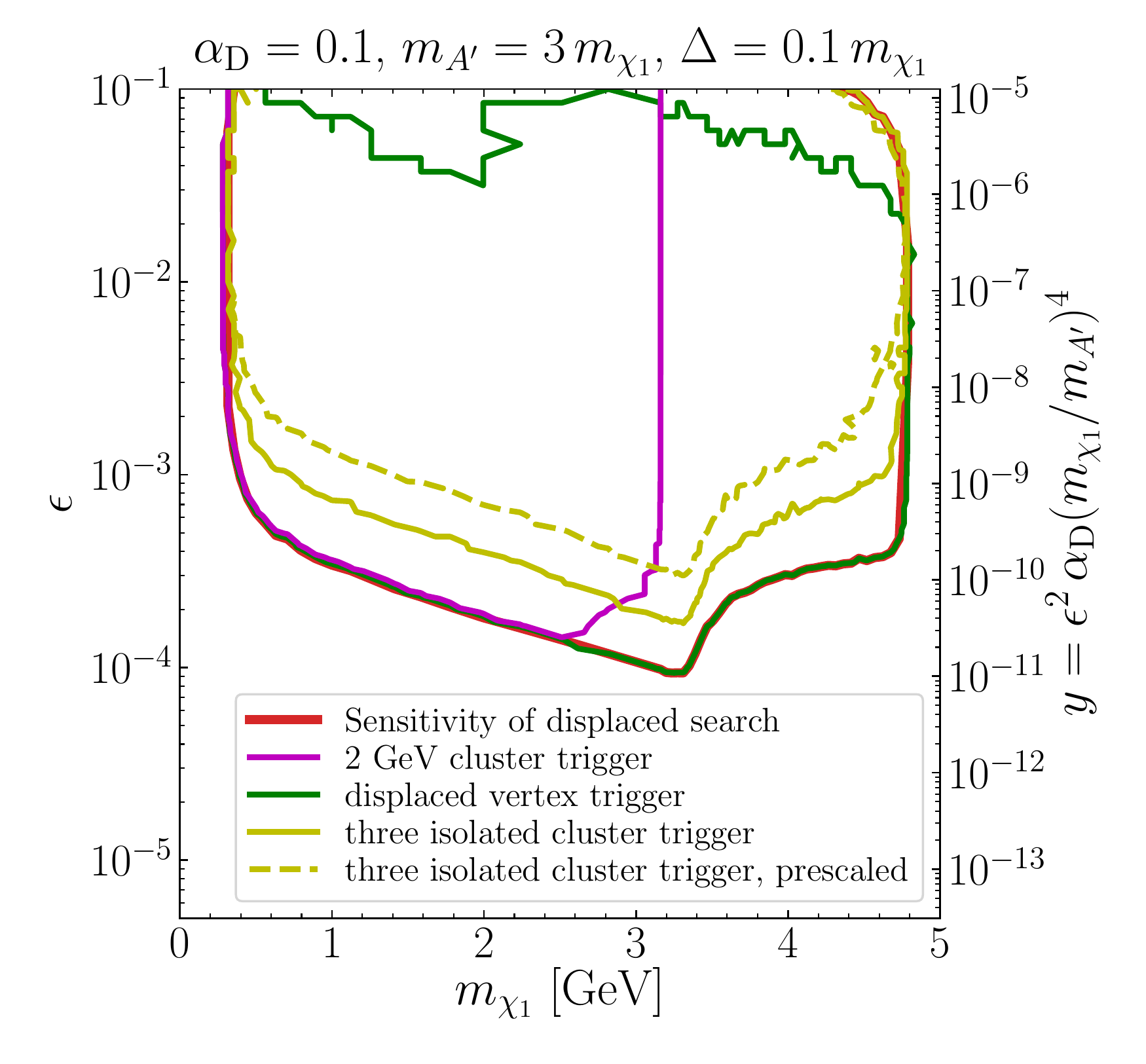}
  \caption{
    Sensitivity of the displaced search, overlaid with the regions where the various triggers are efficient for an integrated luminosity of 50~ab$^{-1}$  for $m_{A^\prime} = 3.0\, m_{\chi_1}, \Delta=0.1 m_{\chi_1}$.
    \label{fig:sensitivity3p0_trigger_highlumi}
  }
\end{figure}

Figure~\ref{fig:sensitivity3p0_trigger} shows the expected sensitivity for the 2\,GeV cluster trigger, the three isolated clusters trigger, and the displaced vertex trigger separately for an integrated luminosity of 20~fb$^{-1}$.
For the smallest values of $\Delta$ the three isolated clusters trigger is inefficient, but it extends the sensitivity significantly towards higher masses for larger $\Delta$.
The displaced vertex trigger has the best sensitivity for large values of $m_{\chi_1}$ and small $\epsilon$, whereas the three isolated clusters trigger adds additional sensitivity for large $\epsilon$.
We note that the rather high $p_T$ and large opening angle requirement make the two-track trigger inefficient. Since the trigger rates of the three isolated clusters trigger are expected to be too high to sustain this trigger at the ultimate luminosities, we investigate the effects of a factor 10 prescale, i.e. randomly dropping nine out of ten events kept by this trigger. Figure~\ref{fig:sensitivity3p0_trigger_highlumi} shows the expected sensitivity for the different triggers for an integrated luminosity of 50~ab$^{-1}$. The sensitivity loss due to this prescale at large values of $\epsilon$ is negligible.

\begin{figure}[tb]
  \centering
  \includegraphics[width=0.99\textwidth]{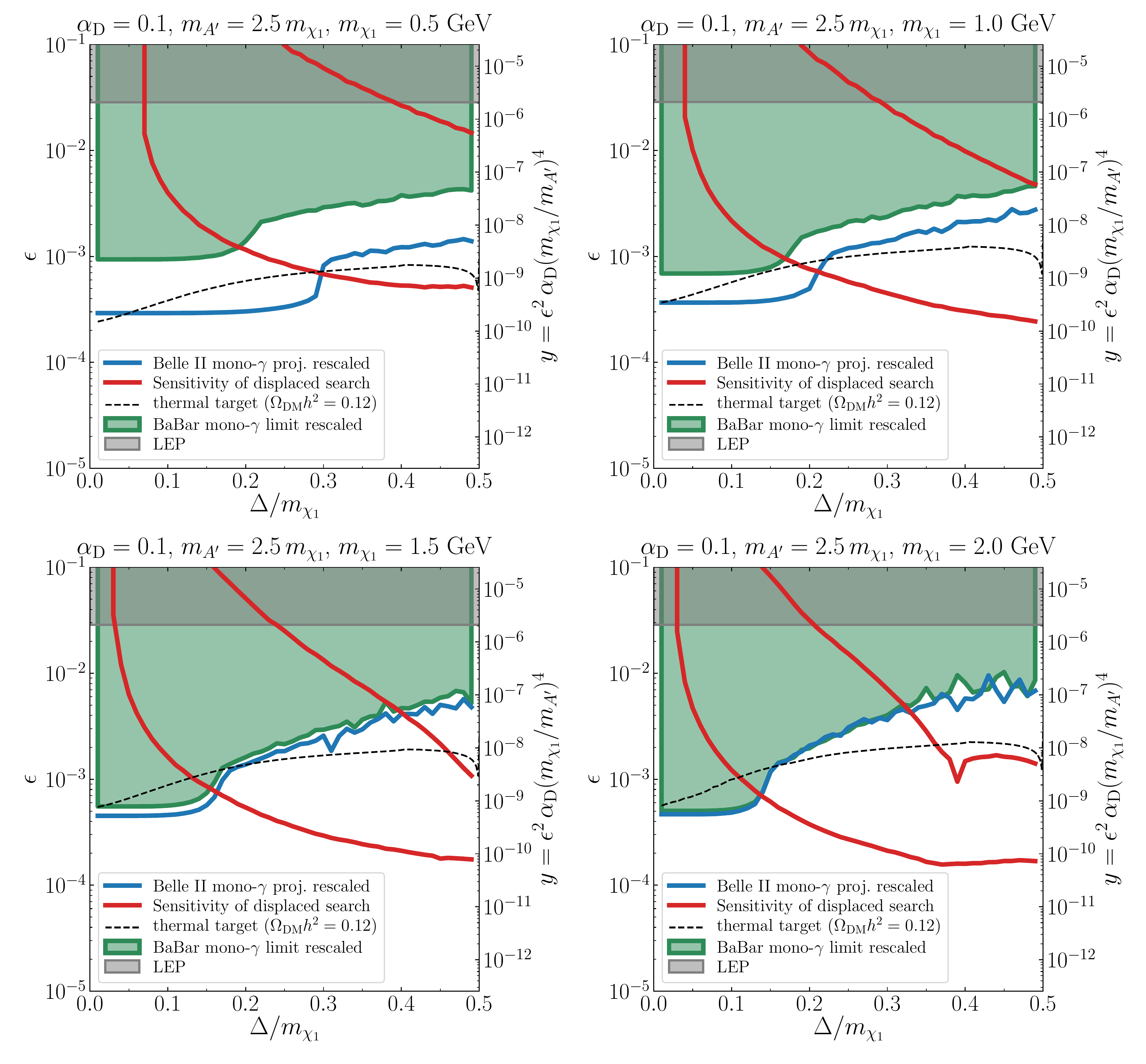}
  \caption{
    Sensitivity of \belletwo to the parameter space of inelastic DM as a function of $\Delta$ for an integrated luminosity of 20 fb$^{-1}$ for $m_{A^\prime} = 2.5\, m_{\chi_1}$. 
    \label{fig:sensitivityDelta}
  }
\end{figure}

Finally, we present our results in a different form in figure~\ref{fig:sensitivityDelta}. Here the mass splitting $\Delta$ is varied explicitly, while the value of $m_{\chi_1}$ is fixed to a different value in each panel. As in figure~\ref{fig:sensitivity2p5} the mass ratio is set to $m_{A^\prime} = 2.5\, m_{\chi_1}$. Again, we observe a strong complementarity between the two different searches. The sensitivity of the mono-photon search decreases with increasing mass splitting, while the sensitivity of the displaced decay search improves. Intriguingly, the combination of the two searches will allow to probe the thermal target for a wide range of DM masses and mass splittings. We note, however, that this conclusion is specific to the assumed ratio of $m_{\chi_1}$ and $m_{A^\prime}$. For $m_{A^\prime} = 3\, m_{\chi_1}$, for example, the thermal target is already partially excluded by the constraint from \babar (see figure~\ref{fig:sensitivity3p0}).

\section{Summary and discussion}
\label{SEC:discussion}

The focus of the present work has been on the phenomenology of dark sectors that contain unstable but long-lived particles. An appealing example for such a dark sector are models of inelastic DM, in which a mass splitting $\Delta$ between two dark sector states $\chi_1$ and $\chi_2$ ensures that constraints from the CMB and direct detection experiments are evaded. The heavier state $\chi_2$ can have a decay length comparable to the typical size of particle physics experiments, making this model an interesting benchmark for searches for displaced vertices.

We have investigated the sensitivity of \belletwo for the key signature of this model: a lepton pair originating from a displaced vertex in association with a single photon. We have identified the most sensitive detector regions and determined selection cuts that suppress the relevant backgrounds to a negligible level. We have furthermore calculated the sensitivity of mono-photon searches at \belletwo and \babar by determining the probability that $\chi_2$ escapes from the detector before decaying or that the decay products are too soft to be observed.

\begin{figure}[tb]
  \centering
  \includegraphics[width=\linewidth]{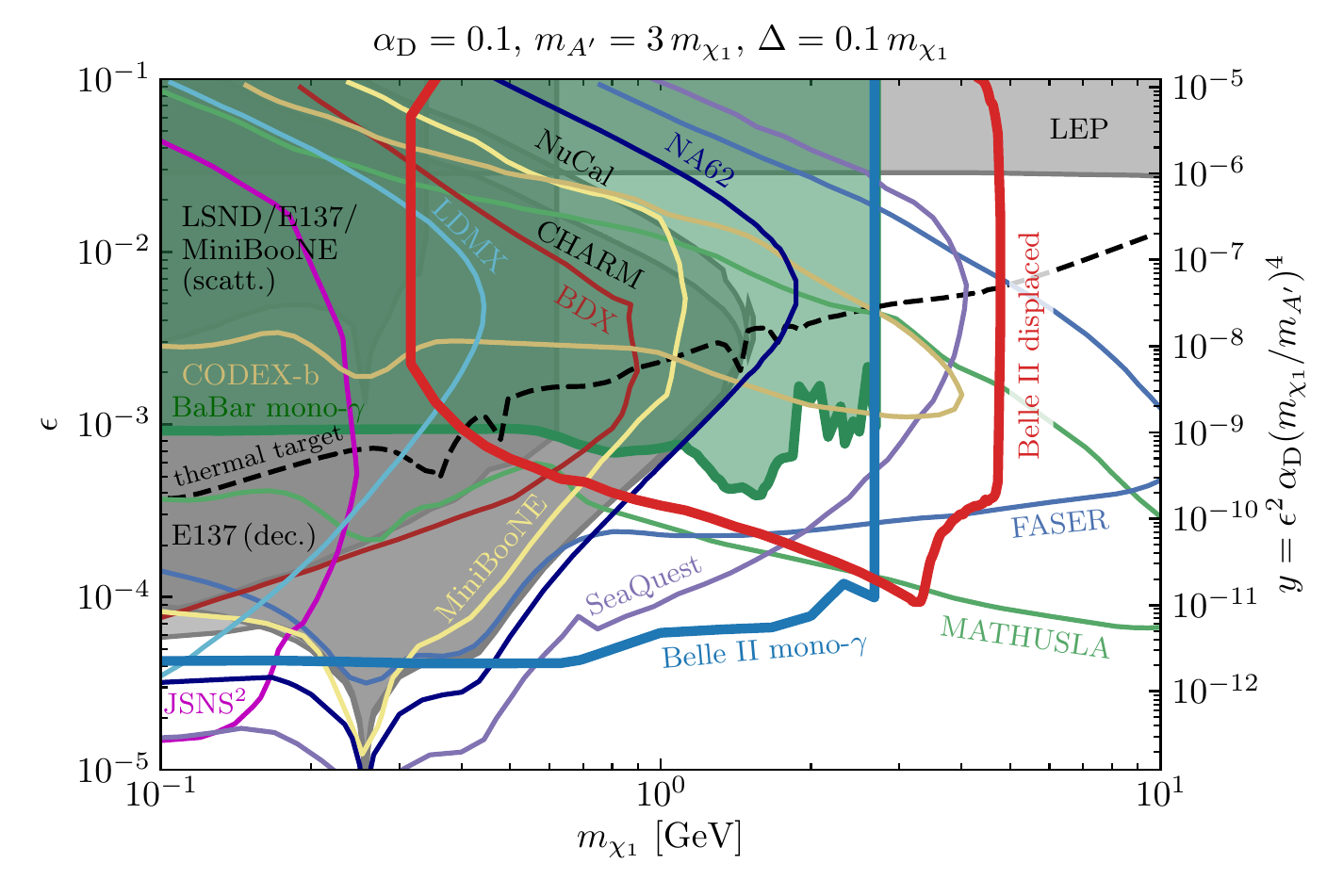}
  \caption{
    Comparison of our results with various other experiments. The \belletwo results are given for $\mathcal{L} = \unit[50]{ab}^{-1}$: for the displaced search, the number of events is calculated with $\mathcal{L} = \unit[50]{ab}^{-1}$, i.e., assuming that this search is still background-free. For the mono-photon search, we rescale the previously found sensitivity with the $\sqrt[4]{\mathcal{L}}$. 
    \label{fig:sensitivity_comparison}
  }
\end{figure} 

Of course, \belletwo is not the only experiment promising to probe deeper into the parameter space of inelastic DM. In figure~\ref{fig:sensitivity_comparison} we show a comparison of the ultimate reach of \belletwo (assuming an integrated luminosity of $50\,\mathrm{ab^{-1}}$) with the projected sensitivities of various proposed experiments to search for long-lived particles. Note that most of the projections shown in figure~\ref{fig:sensitivity_comparison} stem from experiments that are still in early stages of their development. \belletwo in contrast is already taking data and should be able to provide first results within the next few years. 

We emphasize that we assume $\Delta = 0.1 \, m_{\chi_1}$ in figure~\ref{fig:sensitivity_comparison} simply because this choice is commonly used in the literature for sensitivity estimates. The sensitivity of \belletwo for different values of $\Delta / m_{\chi_1}$ are provided in figure~\ref{fig:sensitivity3p0} for an integrated luminosity of $20\,\mathrm{fb^{-1}}$. For larger ratios $\Delta / m_{\chi_1}$, additional decay modes like $\chi_2 \to \chi_1 + \text{hadrons}$ become important and the decay length of $\chi_2$ decreases rapidly. In this case the sensitivity of experiments like FASER (which requires a decay length of about $500\,\mathrm{m}$ in the laboratory frame) are strongly suppressed, while the displaced decay search at \belletwo remains sensitive even for decay lengths below $1\,\mathrm{cm}$. Moreover, the two different signatures discussed in the present work are highly complementary in the sense that the mono-photon search is most sensitive for small $\Delta$, while the displaced vertex search performs best for large $\Delta$ (see figure~\ref{fig:sensitivityDelta}).

As part of this work we have also provided an improved calculation of the thermal target for inelastic DM, which is indicated by the black dashed line in figure~\ref{fig:sensitivity_comparison}. For the specific parameter combination chosen in this figure, large parts of the thermal target are already excluded by the mono-photon bound from BaBar. However, we have shown that this conclusion depends sensitively on the ratio of the DM mass and the dark photon mass (see figure~\ref{fig:relic}) and that for example for $m_{A'} = 2.5 m_{\chi_1}$ the thermal target is essentially not probed by existing constraints (see figure~\ref{fig:sensitivity2p5}).

Finally, we point out that the sensitivity of the displaced vertex search at \belletwo relies crucially on the implementation of suitable triggers. We have identified a number of existing triggers that can in principle be used to search for displaced lepton pairs, but the trigger rate may be too high to make use of the full data set. There is hence clear need for the development of a dedicated displaced vertex trigger. By fully exploiting the potential of such a trigger, \belletwo may soon join the growing number of experiments searching for hidden sectors with long-lived particles. The specific combination of centre-of-mass energy and detector geometry makes \belletwo complementary to other proposals and offers a unique opportunity to explore uncharted territory.

\acknowledgments

We thank Elias Bernreuther and Michael Kr{\"a}mer for useful discussions. We also thank Yu-Dai Tsai and Patrick deNiverville for helpful correspondence and for sharing their data. This work is funded by the Deutsche Forschungsgemeinschaft (DFG) through Germany's Excellence Strategy -- EXC 2121 ``Quantum Universe'' -- 390833306, the Emmy Noether Grant No.\ KA 4662/1-1, and the Collaborative Research Center TRR 257 ``Particle Physics Phenomenology after the Higgs Discovery'', as well as by the ERC Starting Grant ``NewAve'' (638528), the Helmholtz (HGF) Young Investigators Grant No.\ VH-NG-1303, the Science and Technology Facilities Council~(STFC) grant ST/P000770/1 and NSERC~(Canada).

\providecommand{\href}[2]{#2}\begingroup\raggedright\endgroup

\end{document}